\documentclass{aa}  

\usepackage{graphicx}
\usepackage{txfonts}

\usepackage{subcaption}
\usepackage{rotating}
\usepackage{multirow}
\usepackage{arydshln}
\usepackage{tikz}

\DeclareMathOperator{\argmin}{argmin}

\DeclareMathOperator{\Real}{Re}
\DeclareMathOperator{\Imag}{Im}
\DeclareMathOperator{\diag}{diag}
\DeclareMathOperator{\Id}{Id}

\newcommand{\mathC}{\mathbb{C}}
\newcommand{\mathE}{\mathbb{E}}

\newcommand{\mathN}{\mathbb{N}}
\newcommand{\mathP}{\mathbb{P}}

\newcommand{\mathR}{\mathbb{R}}

\newcommand{\MC}{\mathcal C}

\newcommand{\MI}{\mathcal I}

\newcommand{\MM}{\mathcal M}
\newcommand{\MN}{\mathcal N}

\newcommand{\bb}{\boldsymbol b}

\newcommand{\bn}{\boldsymbol n}

\newcommand{\br}{\boldsymbol r}

\newcommand{\BBA}{\mathbf A}
\newcommand{\BBB}{\mathbf B}

\newcommand{\BBF}{\mathbf F}
\newcommand{\BBI}{\mathbf I}

\newcommand{\BBJ}{\mathbf J}

\newcommand{\BBP}{\mathbf P}

\newcommand{\BBS}{\mathbf S}

\newcommand{\rmd}{\mathrm{d}}

\newcommand{\SFK}{\mathsf{K}}

\newcommand{\SFN}{\mathsf{N}}

\newcommand{\SFX}{\mathsf{X}}
\newcommand{\SFY}{\mathsf{Y}}

\newcommand{\SFLambda}{\mathsf{\Lambda}}
\newcommand{\SFGamma}{\mathsf{\Gamma}}

\newcommand{\bbeta}{\boldsymbol \beta}
\newcommand{\bgamma}{\boldsymbol \gamma}

\newcommand{\bepsilon}{\boldsymbol \epsilon}

\newcommand{\bkappa}{\boldsymbol \kappa}

\newcommand{\bnu}{\boldsymbol \nu}

\newcommand{\btheta}{\boldsymbol \theta}

\newcommand{\blambda}{\boldsymbol \lambda}

\newcommand{\BSigma}{\boldsymbol \Sigma}

\newcommand{\bzero}{\boldsymbol 0}

\newcommand{\iif}{\;\Longleftrightarrow\;}

\newcommand{\red}[1]{\textcolor{red}{#1}}

\newcommand{\range}[2]{\left\{#1, \ldots, #2\right\}}
\newcommand{\smallrange}[2]{\{#1, \ldots, #2\}}

\newcommand{\oneto}[1]{\range{1}{#1}}
\newcommand{\smalloneto}[1]{\smallrange{1}{#1}}

\newcommand{\pair}[2]{\left(#1,\, #2\right)}

\newcommand{\set}[2]{\left\{ #1 : #2 \right\}}

\newcommand{\Bigset}[2]{\Bigl\{ #1 : #2 \Bigr\}}

\newcommand{\interval}[2]{\left[#1,\, #2\right]}

\newcommand{\intervalexcllr}[2]{\left]#1,\, #2\right[}

\newcommand{\biginterval}[2]{\bigl[#1,\, #2\bigr]}

\newcommand{\zeroone}{\interval{0}{1}}

\newcommand{\zeroexcloneexcl}{\intervalexcllr{0}{1}}

\newcommand{\tuple}[2]{\{#1,\, #2\}}

\newcommand{\Proba}{\mathP}
\newcommand{\Expval}{\mathE}

\newcommand{\BigcondprobaPrior}[3]{\Proba_{#3}\Bigl\{\, #1 \bigm| #2 \,\Bigr\}}

\newcommand{\Bigcondexpval}[2]{\Expval\Bigl[ #1 \bigm| #2 \Bigr]}

\newcommand{\bigcondexpvalPrior}[3]{\Expval_{#1}\bigl[ #2 \,|\, #3 \bigr]}
\newcommand{\BigcondexpvalPrior}[3]{\Expval_{#1}\Bigl[ #2 \bigm| #3 \Bigr]}

\renewcommand{\texttt}[1]{%
	\begingroup
	\ttfamily
	\begingroup\lccode`~=`/\lowercase{\endgroup\def~}{/\discretionary{}{}{}}%
	\begingroup\lccode`~=`[\lowercase{\endgroup\def~}{[\discretionary{}{}{}}%
	\begingroup\lccode`~=`.\lowercase{\endgroup\def~}{.\discretionary{}{}{}}%
	\catcode`/=\active\catcode`[=\active\catcode`.=\active
	\scantokens{#1\noexpand}%
	\endgroup
} 

\newcommand{\norm}[1]{\left\|#1\right\|}

\newcommand{\qand}{\quad\mbox{and}\quad}
\newcommand{\qqand}{\qquad\mbox{and}\qquad}

\newcommand{\qqwith}{\qquad\mbox{with}\qquad}

\newcommand{\todo}[1][]{%
  	\red{\ifthenelse{\equal{#1}{}}%
    	{[TODO]}%
    	{[TODO: #1]}%
  	}\xspace
}

\makeatletter
\DeclareRobustCommand\onedot{\futurelet\@let@token\@onedot}
\def\@onedot{\ifx\@let@token.\else.\null\fi\xspace}

\def\eg{\emph{e.g}\onedot} 
\def\ie{\emph{i.e}\onedot}

\def\iid{i.i.d\onedot} 

\makeatother

\DeclareMathOperator*{\card}{card}

\DeclareMathOperator*{\herm}{\ast}
\DeclareMathOperator*{\ellip}{ell}

\DeclareMathOperator*{\ks}{ks}
\DeclareMathOperator*{\wiener}{wien}
\DeclareMathOperator*{\mcalens}{mcal}

\DeclareMathOperator*{\rc}{rc}
\DeclareMathOperator*{\cqr}{cqr}

\DeclareMathOperator*{\gauss}{G}
\DeclareMathOperator*{\sparse}{S}

\newcommand{\reducedshear}{g}
\newcommand{\shearmap}{\bgamma}
\newcommand{\convmap}{\bkappa}
\newcommand{\ellmap}{\bepsilon}
\newcommand{\noise}{\bn}
\newcommand{\residual}{\br}

\newcommand{\stdval}{\sigma}

\newcommand{\shearmapRand}{\boldsymbol{\SFGamma}}
\newcommand{\convmapRand}{\boldsymbol{\SFK}}
\newcommand{\noiseRand}{\boldsymbol{\SFN}}

\newcommand{\covmatr}{\BSigma}
\newcommand{\covmatrNoise}{\covmatr_{\noise}}
\newcommand{\covmatrConv}{\covmatr_{\convmap}}
\newcommand{\covmatrRec}{\covmatr_{\convmapEstimate}}

\newcommand{\fouriercovmatr}{\BBP}
\newcommand{\fouriercovmatrConv}{\fouriercovmatr_{\convmap}}
\newcommand{\fouriermatr}{\BBF}
\newcommand{\fouriermatrHerm}{\fouriermatr^{\herm}}

\newcommand{\convToShear}{\BBA}
\newcommand{\shearEstToConv}{\BBB}
\newcommand{\convToShearPseudoinv}{\convToShear^{\!\dagger}}

\newcommand{\convmapEstimate}{\hat\convmap}
\newcommand{\convmapEstimateLow}{\convmapEstimate^{-}}
\newcommand{\convmapEstimateHigh}{\convmapEstimate^{+}}
\newcommand{\convmapEstimateLowCalibparam}{\convmapEstimateLow_{\calibparam}}
\newcommand{\convmapEstimateHighCalibparam}{\convmapEstimateHigh_{\calibparam}}
\newcommand{\convmapEstimateLowCalibparamRCPS}{\convmapEstimateLow_{\calibparamRCPS{\alpha}{\delta}}}
\newcommand{\convmapEstimateHighCalibparamRCPS}{\convmapEstimateHigh_{\calibparamRCPS{\alpha}{\delta}}}

\newcommand{\convmapEstimateKs}{\convmapEstimate_{\ks}}

\newcommand{\convmapEstimateLowRCPS}{\convmapEstimateLow_{\rc}}
\newcommand{\convmapEstimateHighRCPS}{\convmapEstimateHigh_{\rc}}
\newcommand{\convmapEstimateLowCQR}{\convmapEstimateLow_{\cqr}}
\newcommand{\convmapEstimateHighCQR}{\convmapEstimateHigh_{\cqr}}
\newcommand{\residualEstimate}{\hat\residual}

\newcommand{\convmapEstimateRand}{\hat\convmapRand}
\newcommand{\convmapEstimateLowRand}{\convmapEstimateRand^{-}}
\newcommand{\convmapEstimateHighRand}{\convmapEstimateRand^{+}}
\newcommand{\convmapEstimateLowCalibparamRand}{\convmapEstimateLowRand_{\calibparam}}
\newcommand{\convmapEstimateHighCalibparamRand}{\convmapEstimateHighRand_{\calibparam}}
\newcommand{\convmapEstimateLowCalibparamRCPSRand}{\convmapEstimateLowRand_{\calibparamRCPSRand{\alpha}{\delta}}}
\newcommand{\convmapEstimateHighCalibparamRCPSRand}{\convmapEstimateHighRand_{\calibparamRCPSRand{\alpha}{\delta}}}
\newcommand{\convmapEstimateLowRCPSRand}{\convmapEstimateLowRand_{\rc}}
\newcommand{\convmapEstimateHighRCPSRand}{\convmapEstimateHighRand_{\rc}}
\newcommand{\convmapEstimateLowCQRRand}{\convmapEstimateLowRand_{\cqr}}
\newcommand{\convmapEstimateHighCQRRand}{\convmapEstimateHighRand_{\cqr}}

\newcommand{\riskfun}{R}
\newcommand{\riskfunUpperestimate}[1]{\riskfun_{#1}^+}
\newcommand{\calibfun}{g}

\newcommand{\imgsize}{K}
\newcommand{\ngalperpix}[1]{N_{#1}}
\newcommand{\sizeCalibrationset}{n}
\newcommand{\sizeCalibTestset}{m}

\newcommand{\calibparam}{\lambda}
\newcommand{\calibparamVec}{\blambda}
\newcommand{\calibparamRand}{\SFLambda}
\newcommand{\calibparamVecRand}{\boldsymbol{\SFLambda}}

\newcommand{\calibparamCQR}[1]{\calibparam^{(#1)}}
\newcommand{\calibparamVecCQR}[1]{\calibparamVec^{(#1)}}
\newcommand{\calibparamRCPS}[2]{\calibparam^{(#1,\, #2)}}

\newcommand{\calibparamCQRRand}[1]{\calibparamRand^{(#1)}}
\newcommand{\calibparamVecCQRRand}[1]{\calibparamVecRand^{(#1)}}
\newcommand{\calibparamRCPSRand}[2]{\calibparamRand^{(#1,\, #2)}}

\newcommand{\inputOutput}{(x,\, y)}
\newcommand{\calibrationset}[1]{(x_i,\, y_i)_{i = 1}^{#1}}
\newcommand{\datasetMM}[2]{(\shearmap_i,\, \convmap_i)_{i = #1}^{#2}}
\newcommand{\datasetMMpix}[3]{(\shearmap_i,\, \convmap_i[#3])_{i = #1}^{#2}}
\newcommand{\calibrationsetMM}{\datasetMM{1}{\sizeCalibrationset}}
\newcommand{\testsetMM}{\datasetMM{\sizeCalibrationset + 1}{\sizeCalibTestset}}
\newcommand{\calibrationsetMMpix}[1]{\datasetMMpix{1}{\sizeCalibrationset}{#1}}

\newcommand{\hoeffding}{\riskfunUpperestimate{\delta,\, \calibparam}\left(
    \calibrationsetMM
\right)}
\newcommand{\bighoeffding}{\riskfunUpperestimate{\delta,\, \calibparam}\bigl(
    \calibrationsetMM
\bigr)}
\newcommand{\hoeffdingRand}{\riskfunUpperestimate{\delta,\, \calibparam}\left(
    \calibrationsetMMRand
\right)}
\newcommand{\riskfunCalibparam}{\riskfun_\calibparam}

\newcommand{\riskfunCalibparamRCPSRand}{
    \riskfun_{\calibparamRCPSRand{\alpha}{\delta}}
}

\newcommand{\inputOutputRand}{(\SFX,\, \SFY)}
\newcommand{\calibrationsetRand}[1]{(\SFX_i,\, \SFY_i)_{i = 1}^{#1}}
\newcommand{\datasetMMRand}[2]{(\shearmapRand_i,\, \convmapRand_i)_{i = #1}^{#2}}

\newcommand{\calibrationsetMMRand}{\datasetMMRand{1}{\sizeCalibrationset}}

\newcommand{\predinterv}{\hat\MC}
\newcommand{\predintervCalib}[1]{\predinterv_{#1}}

\newcommand{\pdf}[2]{f_{#1}\!\left(#2\right)}
\newcommand{\condpdf}[3]{f_{#1}\!\left(#2 \,|\, #3\right)}
\newcommand{\cond}[2]{#1 \,|\, #2}

\newtheorem{hypothesis}{Hypothesis}{\bfseries}{\itshape}

\begin{document}

\title{Distribution-free uncertainty quantification for inverse problems: application to weak lensing mass mapping}


    \author{
            H. Leterme\inst{1,2}
        \and
            J. Fadili\inst{1}
        \and J.-L. Starck\inst{2,3}
    }

   \institute{Université Caen Normandie, ENSICAEN, CNRS, Normandie Univ, GREYC UMR 6072, F-14000 Caen, France\\
              \email{hubert.leterme@ensicaen.fr}
        \and
          Universit\'e Paris-Saclay, Universit\'e Paris Cit\'e, CEA, CNRS, AIM, 91191, Gif-sur-Yvette, France
         \and
             Institutes of Computer Science and Astrophysics, Foundation for Research and Technology Hellas (FORTH), Greece       
             }
             

 
\abstract
{}
{   
    In inverse problems, distribution-free uncertainty quantification (UQ) aims to obtain error bars in the reconstruction with coverage guarantees that are independent of any prior assumptions about the data distribution. This allows for a better understanding of how intermediate errors propagate through the pipeline.
    In the context of mass mapping, uncertainties could lead to errors that affects our understanding of the underlying mass distribution, or could propagate to cosmological parameter estimation, thereby impacting the precision and reliability of cosmological models.
    Current surveys, such as Euclid or Rubin, will provide new weak lensing datasets of very high quality. 
    Accurately quantifying uncertainties in mass maps is therefore critical to fully exploit their scientific potential and to perform reliable cosmological parameter inference.


}
{
    In this paper, we extend the conformalized quantile regression (CQR) algorithm, initially proposed for scalar regression, to inverse problems. We compare our approach with another distribution-free approach based on risk-controlling prediction sets (RCPS).
    Both methods are based on a calibration dataset, and offer finite-sample coverage guarantees that are independent of the data distribution. Furthermore, they are applicable to any mass mapping method, including blackbox predictors. In our experiments, we apply UQ on three mass-mapping method: the Kaiser-Squires inversion, iterative Wiener filtering, and the MCALens algorithm.
}
{
    Our experiments reveal that RCPS tends to produce overconservative confidence bounds with small calibration sets, whereas CQR is designed to avoid this issue.
    Although the expected miscoverage rate is guaranteed to stay below a user-prescribed threshold regardless of the mass mapping method, selecting an appropriate reconstruction algorithm remains crucial for obtaining accurate estimates, especially around peak-like structures, which are particularly important for inferring cosmological parameters. Additionally, the choice of mass mapping method influences the size of the error bars.
}
{}

\keywords{cosmology: observations -- methods: statistical -- gravitational lensing: weak}


\maketitle

%


\section{Introduction}
\label{sec:intro}

Mapping the distribution of matter across the universe is essential for advancing our understanding of its evolution and constraining cosmological parameters. While dark matter constitutes approximately 85\% of the observable universe's total mass according to the $\Lambda$CDM model, it cannot be directly observed. Instead, its presence can be inferred from gravitational effects, such as the deflection of light rays from distant galaxies. In the weak lensing regime, this phenomenon causes anisotropic stretching of galaxy images, known as shear, which can be used to reconstruct mass maps \citep[for a detailed review on the topic, see][]{Kilbinger2015}.
Estimating shear involves detecting statistical anomalies in galaxy ellipticities \citep{Kaiser1993}. This task is complicated by inherent noise, making weak lensing mass mapping an ill-posed inverse problem without appropriate priors on the matter density field. Furthermore, missing data often occur due to bright objects in the foreground, all the more compromising straightforward solutions. Recent methodologies, including some relying on deep learning, have aimed to enhance mass map reconstructions. However, quantifying uncertainty, particularly with blackbox predictors, remains a significant challenge.

Having accurate uncertainty estimates in inverse problems is crucial for obtaining reliable solutions and 
contributing to a deeper scientific understanding of our data. In the case of mass mapping, the main motivations to derive uncertainty quantifiers are: i) uncertainties could lead to errors, impacting our analysis of the underlying mass distribution and affecting theories on the nature of dark matter; ii) mass maps are also used for cosmological parameter inference using high-order statistics, where any error could propagate, leading to uncertainties in these parameters; iii) uncertainties may mask physical models of galaxy and cluster growth, affecting our understanding of the processes that determine the evolution of these structures over cosmic time; and iv) uncertainties could introduce bias in the measurements of cluster masses. Therefore, it is essential to not only improve the quality of the mass reconstructions but also to quantify and minimize uncertainties to optimize the potential of future surveys such as Euclid or Rubin.

In this paper, we build on a paradigm called conformal prediction \citep{Vovk2005,Lei2014} in the context of weak lensing mass mapping, using simulated data. Specifically, we propose a novel extension of the conformalized quantile regression (CQR) algorithm \citep{Romano2019} to inverse problems, especially mass mapping. Relying on a calibration set, it is applicable to any prediction method, including blackbox deep learning models, and offers distribution-free, per-pixel coverage guarantees with prescribed confidence levels.
For the sake of comparison, we also applied a calibration procedure based on risk-controlling prediction sets (RCPS) \citep{Angelopoulos2022b} to the problem at hand. We observed that the latter approach tends to produce overconservative confidence bounds with small calibration sets, unlike CQR.
We tested the two calibration methods on three mass mapping techniques: the Kaiser-Squires inversion \citep{Kaiser1993}, the forward-backward proximal iterative Wiener filtering \citep{Bobin2012}, and the MCALens iterative algorithm \citep{Starck2021}.
We evaluated the methods in terms of miscoverage rate and prediction interval size.
 

This paper is organized as follows. In Sect.~\ref{sec:sota}, we present a state-of-the-art review on the weak lensing inverse mass mapping problem and the currently used quantification methods. Then, Sect.~\ref{sec:dfuq} introduces and compares the two distribution-free calibration approaches (CQR and RCPS). In Sect.~\ref{sec:exp}, we describe our experimental settings and present the results, followed by a discussion in Sect.~\ref{sec:discussion}. Finally, Sect.~\ref{sec:concl} concludes the paper.

\section{State of the art on the mass mapping inverse problem with uncertainty quantification}
\label{sec:sota}

Throughout the paper, we use the following notation conventions. Deterministic vectors are denoted with bold lower-case Greek or Latin letters ($\convmap$), while random vectors are represented with bold sans-serif capital letters ($\convmapRand$). Deterministic matrices are indicated by standard bold capital letters ($\convToShear$). Furthermore, indexing is done using brackets ($\convmap[k]$,  $\convmapRand[k]$ or $\convToShear[k,\, l]$).

The weak lensing mass mapping problem consists in recovering a convergence map $\convmap \in \mathR^{\imgsize^2}$ from an observed, noisy shear map $\shearmap \in \mathC^{\imgsize^2}$. Both fields have been discretized over a square grid of size $\imgsize \times \imgsize$, with $\imgsize \in \mathN$, and are represented as flattened one-dimensional vectors. The relationship between the shear and convergence maps is expressed as follows:
\begin{equation}
    \shearmap = \convToShear \convmap + \noise,
\label{eq:invprob}
\end{equation}
where $\convToShear \in \mathC^{\imgsize^2 \times \imgsize^2}$ is a known linear operator, referred to as the inverse Kaiser-Squires filter, and $\noise \in \mathC^{\imgsize^2}$ is the realization of a Gaussian noise $\noiseRand$ with zero mean and known diagonal covariance matrix $\covmatrNoise \in \mathR^{\imgsize^2 \times \imgsize^2}$. A more detailed description of the problem is provided in Appendix~\ref{sec:appendix_pb}.

\subsection{Existing mass mapping methods}
\label{subsec:sota_mm}

The most simple---and widely used---algorithm for performing mass mapping is the Kaiser-Squires (KS) method \citep{Kaiser1993}, which consists in inverting the linear operator $\convToShear$ in the Fourier space, then applying a Gaussian smoothing to reduce the noise. However, it yields poor results because it does not properly handle noise and missing data. More recent works have proposed to incorporate handcrafted priors in the optimization problem, leading to improved reconstruction. Among those, we can cite an iterative Wiener algorithm based on a Gaussian prior \citep{Bobin2012}, the GLIMPSE2D algorithm based on a sparse prior in the wavelet domain \citep{Lanusse2016}, and the MCALens algorithm \citep{Starck2021}, which considers the combination of a Gaussian component and a sparse component. More details on these methods are given in Appendix~\ref{sec:appendix_mm}.

The above approaches are essentially model-driven. That is, their design is based on knowledge about the underlying physics, and as such requires handcrafted modeling that may be over-simplifying. Alternatively, data-driven approaches, which take advantage of recent breakthroughs in deep learning, rely on data to learn priors and accurately reconstruct mass maps. Among those, we can cite denoising approaches based on adversarial networks \citep{Shirasaki2019,Shirasaki2021}, DeepMass \citep{Jeffrey2020}, which takes as input the Wiener solution and outputs an enhanced point estimate by using a UNet architecture, and DLPosterior \citep{Remy2023}, which draws samples from the full Bayesian posterior distribution.

\subsection{Uncertainty quantification}
\label{subsec:sota_uq}

We consider a point estimate of the convergence map, denoted by $\convmapEstimate \in \mathR^{\imgsize^2}$, obtained using one of the mass mapping method presented in Sect.~\ref{subsec:sota_mm}.
Uncertainty quantification (UQ) involves estimating lower and upper bounds $\convmapEstimateLow$ and $\convmapEstimateHigh$, such that the probability of miscoverage remains below a pre-specified threshold $\alpha \in \zeroexcloneexcl$. The various approaches to achieve this can be categorized into two main types: frequentist (where the ground truth $\convmap$ is a deterministic unknown vector), and Bayesian (where $\convmap$ is the outcome of a random vector $\convmapRand$ associated with a given prior distribution). In the following sections, we review both frameworks, discuss their limitations, and introduce the need for calibration.

\subsubsection{Frequentist framework}
\label{subsubsec:sota_uq_freq}

In this framework, the ground truth convergence map $\convmap$ is deterministic. However, due to noise, the observed shear map $\shearmap$ is the outcome of a random vector:
\begin{equation}
    \shearmapRand := \convToShear\convmap + \noiseRand, \qqwith \noiseRand \sim \MN(\bzero,\, \covmatrNoise).
\label{eq:invprob_randfreq}
\end{equation}
Consequently, the point estimate $\convmapEstimate$ and the bounds $\convmapEstimateLow$ and $\convmapEstimateHigh$, which are computed from $\shearmap$, are also outcomes of random vectors, that we respectively denote by $\convmapEstimateRand$, $\convmapEstimateLowRand$, and $\convmapEstimateHighRand$. In this context, we target the following coverage property:%
\begin{equation}
    \Proba\left\{
        \convmap[k] \notin \interval{\convmapEstimateLowRand[k]}{\convmapEstimateHighRand[k]}
    \right\} \leq \alpha,
\label{eq:uq_freq}
\end{equation}
for any pixel $k \in \smalloneto{\imgsize^2}$.
In this section, we review some methods to achieve this.

\paragraph{Analytical formulation.}

We assume that the point estimate $\convmapEstimate$ is obtained with a linear operator:
\begin{equation}
    \convmapEstimate := \shearEstToConv\shearmap, \quad \mbox{thus,} \quad \convmapEstimateRand := \shearEstToConv\shearmapRand.
\label{eq:linearop}
\end{equation}
As shown in Appendix~\ref{sec:appendix_mm}, this applies to the KS \eqref{eq:ks} and Wiener \eqref{eq:wien} solutions. Plugging \eqref{eq:invprob_randfreq} into \eqref{eq:linearop} yields
\begin{equation}
    \convmapEstimateRand = \shearEstToConv\convToShear\convmap + \shearEstToConv\noiseRand.
\end{equation}
Then, we can easily show that $\convmapEstimateRand$ follows a multivariate Gaussian distribution:
\begin{equation}
    \convmapEstimateRand \sim \MN(\shearEstToConv\convToShear\convmap,\, \covmatrRec), \qqwith \covmatrRec := \shearEstToConv\covmatrNoise\shearEstToConv^{\herm}.
\label{eq:convmapEstimate_gaussianvec}
\end{equation}
Now, we consider the following hypothesis:
\begin{hypothesis}
    \label{hyp:unbiased}
    The estimator $\convmapEstimateRand$ is unbiased: $\shearEstToConv\convToShear\convmap = \convmap$.
\end{hypothesis}
We consider the (deterministic) residual vector $\residualEstimate$ satisfying
\begin{equation}
    \residualEstimate[k] := \Phi_k^{-1} (1 - \alpha / 2) > 0, 
\end{equation}
where $\Phi_k$ denotes the cumulative distribution function (CDF) of a Gaussian distribution with zero mean and variance
\begin{equation}
    \sigma_k^2 := \covmatrRec[k,\, k].
\label{eq:variance_est}
\end{equation}
Then, by setting
\begin{equation}
    \convmapEstimateLowRand := \convmapEstimateRand - \residualEstimate \qand \convmapEstimateHighRand := \convmapEstimateRand + \residualEstimate,
\label{eq:lowerUpperBounds}
\end{equation}
we can prove that, under Hyp.~1, the coverage property \eqref{eq:uq_freq} is satisfied.

On the other hand, GLIMPSE2D and MCALens \eqref{eq:mcalens} are associated with nonlinear operators on the form
\begin{equation}
    \convmapEstimate := \shearEstToConv(\shearmap) \cdot \shearmap,
\end{equation}
where the matrix $\shearEstToConv(\shearmap)$ is characterized by a set of active coefficients in a wavelet dictionary,
which are dependent on the input $\shearmap$. Therefore, $\shearEstToConv(\shearmap)$ is the outcome of a random matrix $\shearEstToConv(\shearmapRand)$, which we assume to be noise-insensitive, essentially depending on the true convergence map $\convmap$:
\begin{hypothesis}
    \label{hyp:back2linear}
    $\shearEstToConv(\shearmapRand) = \shearEstToConv(\convToShear\convmap)$.
\end{hypothesis}
Under Hyp.~2, $\shearEstToConv(\shearmapRand)$ is approximated with a deterministic matrix $\shearEstToConv(\convToShear\convmap)$, allowing us to adopt a similar approach as in the linear case \eqref{eq:linearop}, using $\shearEstToConv := \shearEstToConv(\convToShear\convmap)$.

\paragraph{Practical implementations.}

Estimating confidence bounds in the above framework only requires to compute the diagonal elements of $\covmatrRec$, as evidenced in \eqref{eq:variance_est}.
However, we need explicit access to all the elements of $\shearEstToConv$, which is infeasible in practice.


In the KS case, by exploiting the spectral properties $\shearEstToConv$, we can show that
\begin{equation}
    \diag(\covmatrRec) = \diag(\covmatrNoise) \ast |\bb|^2,
\label{eq:2dconv}
\end{equation}
where $\bb$ denotes the first column vector of $\shearEstToConv$ and $\ast$ denotes the 2D circular convolution product.\footnote{The 2D convolution is performed after having reshaped the vectors to $\imgsize \times \imgsize$ matrices.}
In practice,
the convolution filter $|\bb|^2$ is fast-decaying, and therefore can be cropped to a much smaller size with negligible impact on the result.


In a more general case, the diagonal elements of $\covmatrRec$ cannot simply be obtained with a 2D convolution. Alternatively, they can be estimated with a Monte-Carlo approach, by noticing that $\covmatrRec$ is also the covariance matrix of $\shearEstToConv\noiseRand$. Then, by propagating noise realizations through operator $\shearEstToConv$, and by computing the empirical variance of the outputs at the pixel level, we get an empirical, unbiased estimate of the diagonal elements of $\covmatrRec$.
This method is straightforward for linear operators such as KS and Wiener filters, and has also been used by \citet{Starck2021} for MCALens.


\subsubsection{Bayesian framework}
\label{subsubsec:sota_uq_bayes}

Bayesian UQ pursue a different objective than \eqref{eq:uq_freq}. In this framework, the ground truth convergence map $\convmap$ is the outcome of a random vector $\convmapRand$ with unknown distribution $\mu^\ast$. In this context, the inverse problem \eqref{eq:invprob_randfreq} becomes
\begin{equation}
    \shearmapRand := \convToShear\convmapRand + \noiseRand.
\label{eq:invprob_randbayes}
\end{equation}
We consider a prior distribution $\mu$ estimating $\mu^\ast$, that can be either built from expert knowledge (model-driven approaches), or learned from a training set (data-driven approaches). We assume $\mu$ to be associated with a probability density function $\pdf{\mu}{\convmap}$. Then, given an observation $\shearmap$ drawn from $\shearmapRand$, one can derive a posterior density $\condpdf{\mu}{\convmap}{\shearmap}$ satisfying Bayes' rule:
\begin{equation}
    \condpdf{\mu}{\convmap}{\shearmap} \propto \condpdf{\cond{\shearmapRand}{\convmapRand}}{\shearmap}{\convmap} \, \pdf{\mu}{\convmap},
\label{eq:posteriordensity}
\end{equation}
where the likelihood density $\condpdf{\cond{\shearmapRand}{\convmapRand}}{\shearmap}{\convmap}$ corresponds to a multivariate Gaussian distribution with mean $\convToShear\convmap$ and covariance matrix $\covmatrNoise$, according to \eqref{eq:invprob_randbayes}:
\begin{equation}
    \condpdf{\cond{\shearmapRand}{\convmapRand}}{\shearmap}{\convmap} \propto \exp\left(
        -\frac12 \norm{\shearmap - \convToShear \convmap}_{\covmatrNoise^{-1}}^2
    \right).
\end{equation}
Now, we consider uncertainty bounds $\convmapEstimateLow$ and $\convmapEstimateHigh$ satisfying
\begin{equation}
    \BigcondprobaPrior{
        \convmapRand[k] \notin \biginterval{\convmapEstimateLow[k]}{\convmapEstimateHigh[k]}
    }{\shearmapRand = \shearmap}{\mu} \leq \alpha,
\label{eq:uq_bayes}
\end{equation}
where the miscoverage rate for pixel $k \in \smalloneto{\imgsize^2 - 1}$ is obtained by marginalizing the posterior density over all other pixels:
\begin{multline}
    \BigcondprobaPrior{
        \convmapRand[k] \notin \biginterval{a}{b}
    }{\shearmapRand = \shearmap}{\mu}
    \\
    := 1 - \int_{\mathR^{k-1}}\int_{a}^{b}\int_{\mathR^{\imgsize^2 - k}} \condpdf{\mu}{\convmap'}{\shearmap} \, \rmd \convmap'.
\end{multline}
As in Sect.~\ref{subsubsec:sota_uq_freq}, we denote by $\convmapEstimateLowRand$ and $\convmapEstimateHighRand$ the random variables from which $\convmapEstimateLow$ and $\convmapEstimateHigh$ are drawn, which are dependent on $\shearmapRand$.

In practice, computing the full posterior is intractable. Instead, approximate error bars can be obtained using two different families of methods.


\paragraph{Full posterior sampling.}

There exists a broad literature focusing on sampling high-dimensional posterior distributions, using proximal MCMC algorithms with Langevin dynamics \citep{Pereyra2016,Durmus2018,Cai2018a,Pereyra2020,Laumont2022,McEwen2023,Klatzer2024}, or deep generative models based on neural score matching \citep{Remy2023}, as reviewed in Sect.~\ref{subsec:sota_mm}.
From these samples, we can derive a point estimate $\convmapEstimate$, corresponding to an empirical approximation of the posterior mean:
\begin{equation}
    \convmapEstimate \approx \bigcondexpvalPrior{\mu}{\convmapRand}{\shearmapRand = \shearmap} := \int_{\mathR^{\imgsize^2}} \condpdf{\mu}{\convmap'}{\shearmap} \, \convmap' \, \rmd\convmap',
\label{eq:posteriormean}
\end{equation}
by computing the pixelwise empirical mean,
as well as confidence bounds $\convmapEstimateLow$ and $\convmapEstimateHigh$, by computing the pixelwise $(\alpha / 2)$-th and $(1 - \alpha / 2)$-th empirical quantiles, respectively.

\paragraph{Fast Bayesian UQ.}

MCMC sampling methods offer a detailed representation of the posterior distribution, but are known to be computationally expensive. Alternatively, other approaches can estimate pixelwise error bars orders of magnitude faster without the need for high-dimensional sampling.

For instance, assuming an explicit log-concave prior, concentration inequalities \citep{Pereyra2017} can be used to provide a stable, though somewhat conservative, approximation of the highest-posterior density region from the MAP point estimate. This method has been used to compute Bayesian error bars in the contexts of radio-interferometry \citep{Cai2018} and mass mapping \citep{Price2020}. Recently, \citet{Liaudat2023} developed a data-driven model based on similar principles, implementing a learned prior specifically designed to be log-concave.

Alternatively, \citet{Jeffrey2020a} proposed a direct estimation of lower-dimensional marginal posterior distributions (for instance, per-pixel estimation) that can quantify uncertainty without relying on high-dimensional MCMC sampling.


\subsubsection{Limits of the methods}
\label{subsubsec:sota_uq_limits}

The methods reviewed above provide an initial guess for the confidence bounds $\convmapEstimateLow$ and $\convmapEstimateHigh$, but their accuracy strongly depends on the choice of the reconstruction method and / or the prior data distribution.
Let us review the weaknesses of both frameworks more specifically.

\paragraph{Frequentist framework.}

As explained hereafter, Hyp.~1 generally does not hold. That is, applying the mass mapping method on a noise-free shear map $\convToShear\convmap$ does not necessarily accurately recover $\convmap$.

Regarding the KS filter, we have $\shearEstToConv := \BBS\convToShear$, where $\BBS$ denotes a Gaussian low-pass filter. This smoothing operator induces a bias in the solution. In addition, the Wiener and MCALens solutions can be interpreted as maximum a posteriori (MAP) Bayesian estimates, which are purposely biased:
\begin{align}
    \convmapEstimate
        &\in \argmin_{\convmap'} \Bigl\{
            -\log \condpdf{\mu}{\convmap'}{\shearmap}
        \Bigr\}
\label{eq:mapz} \\
        &= \argmin_{\convmap'} \left\{
            \frac12 \norm{\shearmap - \convToShear \convmap'}_{\covmatrNoise^{-1}}^2 - \log \pdf{\mu}{\convmap'}
        \right\},
\label{eq:map}
\end{align}
where $\pdf{\mu}{\convmap'}$ and $\condpdf{\mu}{\convmap'}{\shearmap}$ respectively denote the prior and posterior densities, as described in Sect.~\ref{subsubsec:sota_uq_bayes}.
The MAP estimator aims to find the most probable solution according to the prior distribution $\mu$, given an observation $\shearmap$. Intuitively, if this prior does not align with the true distribution $\mu^\ast$ from which $\convmapRand$ is drawn, then the estimator is likely to produce a poor reconstruction $\convmapEstimate$ of the ground truth $\convmap$, even in the absence of noise.
For example, it is well known that convergence maps are poorly modeled by Gaussian distributions \citep{Starck2021}. Consequently, the Wiener filter, which assumes a Gaussian prior, tends to blur reconstructed convergence maps around peak-like structures.

\begin{remark}
    Bayesian MAP interpretations of regularized variational problems should be approached with caution.
    In fact, this is a possible but not the only interpretation. This is for instance the case for sparse regularization, as employed in GLIMPSE2D and MCALens. 
    We refer readers to \citet{Starck2013} for a detailed discussion on this topic.
\end{remark}

We identify two additional limitations of the frequentist approach. First, in nonlinear cases such as MCALens, Hyp.~2 is only an approximation. Second, the Monte-Carlo estimation of $\diag(\covmatrRec)$ may introduce sampling errors. To correct these errors without the computational burden of increasing the number of samples, the confidence bounds can be adjusted using bootstrapping. This approach, however, is beyond the scope of this paper. Instead, we focus on distribution-free calibration methods, as described in Sect.~\ref{sec:dfuq}.

For all these reasons, the coverage property for the frequentist framework, stated in \eqref{eq:uq_freq}, is no longer guaranteed. As evidenced in our experiments (see Table~\ref{table:results}), this approach tends to underestimate the size of the error bars.

\paragraph{Bayesian framework.}

If the prior distribution $\mu$, from which the confidence bounds $\convmapEstimateLow$ and $\convmapEstimateHigh$ are estimated, does not align with the true unknown distribution $\mu^\ast$ associated with $\convmapRand$, then the empirical miscoverage rate may be inconsistent with \eqref{eq:uq_bayes}, leading to under- or over-confident predictions.
In particular, the data-driven methods rely on both the quality of the training data and the ability of the model to capture the correct prior.

In Sect.~\ref{sec:dfuq}, we will introduce two post-processing calibration procedures, respectively based on conformalized quantile regression (CQR) and risk-controlling prediction sets (RCPS), adjusting the confidence bounds $\convmapEstimateLow$ and $\convmapEstimateHigh$, obtained in either frequentist or Bayesian frameworks. The goal is to get coverage guarantees that do not suffer from the above limitations. Both methods are distribution-free (that is, they do not require any prior assumption on the data distribution), work for any mass mapping method (including blackbox deep-learning models), and provide valid coverage guarantees in finite samples.

\section{Distribution-free uncertainty quantification}
\label{sec:dfuq}

In Sect.~\ref{subsec:dfuq_gen}, we present the CQR algorithm by \citet{Romano2019}, in a generalized formulation that we established. Originally designed for scalar regression, we propose in Sect.~\ref{subsec:dfuq_cmm} an extension to inverse problems, including mass mapping. Then, in Sect.~\ref{subsec:dfuq_rcps}, we review the RCPS algorithm by \citet{Angelopoulos2022b}. In Sect.~\ref{subsec:dfuq_diff}, we present the major differences between the two approaches. Finally, in Sect.~\ref{subsec:dfuq_diffuncalib}, we explain how the theoretical guarantees provided by CQR and RCPS differ from those targeted by frequentist and Bayesian UQ methods reviewed in Sect.~\ref{subsec:sota_uq}.

\subsection{Conformalized quantile regression: general framework}
\label{subsec:dfuq_gen}

Conformal prediction, used in the context of classification \citep{Sadinle2019,Romano2020,Angelopoulos2022a} and quantile regression (CQR) \citep{Romano2019}, offers finite-sample coverage guarantees with user-prescribed confidence levels.
In this section, we present a generalized version of the CQR algorithm, in which we introduce the concept of calibration function.

Let $\inputOutputRand$ denote a pair of random variables taking values in $\MI \times \mathR$, where $\SFX$ is a scalar or multivariate variable of observables, and $\SFY$ is the response variable.
We now consider a prediction interval $\predinterv: \MI \to 2^\mathR$. It satisfies, for any $x \in \MI$,
\begin{equation}
    \predinterv(x) := \left[
        \hat f(x) - \hat r(x),\, \hat f(x) + \hat r(x)
    \right],
\end{equation}
for some prediction functions $\hat f: \MI \to \mathR$ (producing point estimates), and $\hat r: \MI \to \mathR_+$ (producing residuals). In practice, the bounds can be obtained via quantile regression. In this framework, the predictors $\hat f^- := \hat f - \hat r$ and $\hat f^+ := \hat f + \hat r$ are designed to approximate the $(\alpha/2)$-th and $(1 - \alpha/2)$-th quantiles of $\SFY$, for a given $\alpha \in \zeroexcloneexcl$. However, the following risk of miscoverage:
\begin{equation}
    \Proba\bigl\{
        \SFY \notin \predinterv(\SFX)
    \bigr\} \leq \alpha
\label{eq:cqr_guarantee0}
\end{equation}
is only guaranteed asymptotically, for some specific models, and under regularity conditions \citep{Takeuchi2006,Steinwart2011}.

Now, given a calibration parameter $\calibparam \in \interval{a}{b} \subseteq \overline\mathR$ (with $a$ and $b$ possibly infinite), we introduce a calibrated prediction interval as follows:
\begin{equation}
    \predintervCalib{\calibparam}(x) := \left[
        \hat f(x) - \calibfun_\calibparam\!\left(
            \hat r(x)
        \right),\,  \hat f(x) + \calibfun_\calibparam\!\left(
            \hat r(x)
        \right)
    \right],
\label{eq:calibpredinterv}
\end{equation}
where
\begin{equation}
    \calibfun_\calibparam: \mathR_+ \to \mathR_+
\label{eq:calibfun}
\end{equation}
denotes a family of non-decreasing calibration functions parameterized by $\calibparam$, such that $\calibfun_{\calibparam_0} = \Id$ for some specific value $\calibparam_0$ (no calibration), and producing larger prediction intervals with increasing values of $\calibparam$:
\begin{equation}
    \calibparam \leq \calibparam' \implies \forall r \in \mathR_+,\, \calibfun_\calibparam(r) \leq \calibfun_{\calibparam'}(r).
\label{eq:calibfun_nondecreasing}
\end{equation}
Therefore, $\calibfun_\calibparam$ shrinks the initial prediction interval if $\calibparam < \calibparam_0$, and expands it if $\calibparam > \calibparam_0$. Moreover, we assume that $\calibparam \mapsto \calibfun_\calibparam(r)$ is continuous. Examples of such calibration functions are provided in Fig.~\ref{fig:calibfun}.
For instance, in the original paper by \citet{Romano2019}, $\calibfun_\calibparam(r) = \max(r + \calibparam,\, 0)$ for any $\calibparam \in \mathR$ (see Fig.~\ref{subfig:calibfun_1}), and therefore,
\begin{equation}
    \predintervCalib{\calibparam}(x) := \left[
        \hat f^-(x) - \calibparam,\,  \hat f^+(x) + \calibparam
    \right].
\end{equation}

The goal is then to find the smallest value of $\calibparam$ such that \eqref{eq:cqr_guarantee0} is guaranteed with $\predinterv := \predintervCalib{\calibparam}$. For this, we consider a calibration set $\calibrationset{\sizeCalibrationset}$ drawn from $\sizeCalibrationset \in \mathN$ pairs of random variable $\calibrationsetRand{\sizeCalibrationset}$.
We then compute the following conformity scores, for any $i \in \oneto{\sizeCalibrationset}$ (the smaller the better):
\begin{align}
    \calibparam_i
        :&= \min\left\{
            \calibparam \in \interval{a}{b} :\, y_i \in \predintervCalib{\calibparam}(x_i)
        \right\}
\label{eq:conformityscore1} \\
        &= \min\left\{
            \calibparam \in \interval{a}{b} :\, \calibfun_\calibparam\bigl(\hat r(x_i)\bigr) \geq \left|
                \hat f(x_i) - y_i
            \right|
        \right\},
\label{eq:conformityscore2}
\end{align}
and the conformity score is set to $a$ (resp.\@ $b$) if the condition is always (resp.\@ never) met.
The existence of $\calibparam_i \in \interval{a}{b}$ is guaranteed by the continuity of $\calibparam \mapsto \calibfun_\calibparam(r)$.
In the case where $\calibfun_\calibparam(r) := \max(r + \calibparam,\, 0)$ as in the original paper, we have
\begin{equation}
    \calibparam_i = \left|
        \hat f(x_i) - y_i
    \right| - \hat r(x_i).
\end{equation}
Then, we compute the $(1 - \alpha)(1 + 1/\sizeCalibrationset)$-th empirical quantile of $(\calibparam_i)_{i=1}^{\sizeCalibrationset}$, denoted by $\calibparamCQR{\alpha}$. The random variables from which $\calibparam_i$ and $\calibparamCQR{\alpha}$ are drawn are respectively denoted by $\calibparamRand_i$ and $\calibparamCQRRand{\alpha}$. The latter depends on the calibration set $\calibrationsetRand{\sizeCalibrationset}$, but is independent of $\SFX$ and $\SFY$. The term $(1 + 1/\sizeCalibrationset)$, which accounts for finite-sample correction, imposes a lower bound for the target error rate: $\alpha \geq 1 / (\sizeCalibrationset + 1)$. To ease the flow of reading, we write $\SFX_{\sizeCalibrationset+1} := \SFX$ and $\SFY_{\sizeCalibrationset+1} := \SFY$.

The following proposition, for which a proof is provided in Appendix~\ref{appendix:proofcqr}, is a generalization of Theorem~1 by \citet{Romano2019}.

\begin{proposition}
    \label{prop:cqr}
    If $\calibrationsetRand{\sizeCalibrationset + 1}$ are drawn exchangeably from an arbitrary joint distribution, and if the conformity scores $(\calibparamRand_i)_{i=1}^{\sizeCalibrationset + 1}$ are almost surely distinct, then,
    \begin{equation}
        \alpha - \frac1{\sizeCalibrationset + 1} \leq \Proba\left\{
            \SFY \notin \predintervCalib{\calibparamCQRRand{\alpha}}(\SFX)
        \right\} \leq \alpha.
    \label{eq:cqr_guarantee}
    \end{equation}
\end{proposition}

As an example, \iid random variables are exchangeable. More generally, exchangeability implies identical distribution but not necessarily independence.
The almost-surely-distinct condition on $(\calibparamRand_i)_{i=1}^{\sizeCalibrationset + 1}$ implies that each conformity score is almost surely distinct from $\tuple{a}{b}$, indicating that the predictions can be calibrated.
The result stated in \eqref{eq:cqr_guarantee} includes a lower bound for the probability of miscoverage; therefore, CQR avoids overconservative prediction intervals whenever $\sizeCalibrationset$ is large enough.


\subsection{Conformalized mass mapping}
\label{subsec:dfuq_cmm}

The CQR algorithm was initially designed for scalar regression. In the context of mass mapping, where both inputs and outputs are multidimensional, we propose applying CQR to each output pixel individually. This idea was also exploited in a very recent paper by \citet{Narnhofer2024}, though it was limited to Bayesian error quantification. In contrast, our method is a more straightforward extension of the CQR algorithm to inverse problems, which does not restrict to the Bayesian framework. To the best of our knowledge, this direction has not been explored before.



Similarly to Sect.~\ref{subsec:sota_uq}, we denote by $\convmapEstimateLow$ and $\convmapEstimateHigh$ the initial lower- and upper-confidence bounds, obtained using any mass mapping method and any UQ approach. The random vectors from which they are drawn are written $\convmapEstimateLowRand$, and $\convmapEstimateHighRand$. We also denote by
\begin{equation}
    \convmapEstimate := \frac{\convmapEstimateLow + \convmapEstimateHigh}{2} \qqand \residualEstimate := \frac{\convmapEstimateHigh - \convmapEstimateLow}{2}
\end{equation}
the corresponding point estimate and residual.\footnote{
    This point estimate may differ from the one obtained in \eqref{eq:posteriormean} within the Bayesian framework, because the CQR algorithm presented in this paper requires equally sized lower and upper residuals. Although \citet{Romano2019} proposed an asymmetric version of their algorithm, it tends to increase the length of prediction intervals.
}
We consider a calibration set $\calibrationsetMM$, drawn from \iid random pairs of shear and convergence maps $\calibrationsetMMRand$ following the same distribution as $(\shearmapRand,\, \convmapRand)$. We will explain in Sect.~\ref{subsec:exp_settings} how to get such a calibration set. Then, for each pixel $k \in \smalloneto{\imgsize^2}$, we apply CQR such as described in Sect.~\ref{subsec:dfuq_gen} with $\mathC^{\imgsize^2}$ as input space $\MI$, and
\begin{equation}
\begin{matrix}
    x := \shearmap, & y := \convmap[k], & \calibrationset{\sizeCalibrationset} := \calibrationsetMMpix{k}
\end{matrix}.
\end{equation}
The conformity scores are encoded into $\sizeCalibrationset$ score vectors $\calibparamVec_i \in \mathR^{\imgsize^2}$. Similarly, the resulting calibration parameters are represented by a calibration vector $\calibparamVecCQR{\alpha} \in \mathR^{\imgsize^2}$. We denote by $\convmapEstimateLowCQR$ and $\convmapEstimateHighCQR$ the lower- and upper-bounds obtained by calibrating the initial bounds $\convmapEstimateLow$ and $\convmapEstimateHigh$ pixelwise:
\begin{equation}
    \convmapEstimateLowCQR := \convmapEstimate - \calibfun_{\calibparamVecCQR{\alpha}}(\residualEstimate)
    \qqand
    \convmapEstimateHighCQR := \convmapEstimate + \calibfun_{\calibparamVecCQR{\alpha}}(\residualEstimate),
\end{equation}
where we have defined, for any calibration vector $\calibparamVec \in \mathR^{\imgsize^2}$, any residual $\br \in \mathR^{\imgsize^2}$, and any pixel $k \in \smalloneto{\imgsize^2}$,%
\begin{equation}
    \calibfun_{\calibparamVec}(\br)[k] := \calibfun_{\calibparamVec[k]}(\br[k]),
\end{equation}
for a given scalar calibration function $\calibfun_{\calibparam}$, as introduced in \eqref{eq:calibfun}.

Now, we denote by $\calibparamVecRand_i$, $\calibparamVecCQRRand{\alpha}$, $\convmapEstimateLowCQRRand$, and $\convmapEstimateHighCQRRand$ the random vectors from which $\calibparamVec_i$, $\calibparamVecCQR{\alpha}$, $\convmapEstimateLowCQR$, and $\convmapEstimateHighCQR$ are drawn, respectively.
We assume that the conformity scores $(\calibparamVecRand_i[k])_{i=1}^{\sizeCalibrationset + 1}$ are almost surely distinct, for each pixel $k$.
Then, by averaging over all pixels, \eqref{eq:cqr_guarantee} yields the following result:
\begin{equation}
    \alpha - \frac1{\sizeCalibrationset + 1} \leq \Expval\left[
        L\left(
            \convmapRand,\, \convmapEstimateLowCQRRand,\, \convmapEstimateHighCQRRand
        \right)
    \right] \leq \alpha,
\label{eq:cqr_guarantee1}
\end{equation}
where we have defined the imagewise miscoverage rate:
\begin{equation}
    L(\convmap,\, \convmapEstimateLow,\, \convmapEstimateHigh) :=
    \frac{
        \card\,\Bigset{
            k \in \MM
        }{
            \convmap[k] \notin \bigl[
                \convmapEstimateLow[k],\, \convmapEstimateHigh[k]
            \bigr]
        }
    }{\card \MM},
\label{eq:miscoveragerate}
\end{equation}
where $\MM \subset \smalloneto{\imgsize^2}$ denotes a set of active pixels. More details on masked data are provided in Appendix~\ref{subsec:appendix_pb_mask}.

\subsection{Risk-controlling prediction sets}
\label{subsec:dfuq_rcps}

In this section, we present an alternative calibration approach based on risk-controlling prediction sets (RCPS). Initially developed by \citet{Bates2021} in the context of classification, it was later extended to inverse problems \citep{Angelopoulos2022b} and diffusion models \citep{Horwitz2022,Teneggi2023}. Relying on Hoeffding's inequality \citep{Boucheron2004}, this approach also provides distribution-free, finite-sample coverage guarantees with user-prescribed confidence levels. 
In contrast to CQR, RCPS aims at controlling the risk of statistical anomalies in the calibration set. However, it does not prevent overconservative confidence intervals.

We consider a family of calibration functions $\calibfun_\calibparam: \mathR_+ \to \mathR_+$ as introduced in Sect.~\ref{subsec:dfuq_gen}. Then, for a given parameter $\calibparam \in \mathR$, we consider the adjusted bounds $\convmapEstimateLowCalibparam$ and $\convmapEstimateHighCalibparam$ satisfying
\begin{equation}
    \convmapEstimateLowCalibparam := \convmapEstimate - \calibfun_\calibparam(\residualEstimate)
    \qqand
    \convmapEstimateHighCalibparam := \convmapEstimate + \calibfun_\calibparam(\residualEstimate).
\label{eq:lowerUpperBoundsCalib}
\end{equation}
Considering the corresponding random vectors $\convmapEstimateLowCalibparamRand$ and $\convmapEstimateHighCalibparamRand$, we define the risk $\riskfunCalibparam$ as the expected miscoverage rate:
\begin{equation}
    \riskfunCalibparam := \Expval\left[
            L\left(
                \convmapRand,\, \convmapEstimateLowCalibparamRand,\, \convmapEstimateHighCalibparamRand
            \right)
        \right],
\label{eq:riskfun}
\end{equation}
where $L(\convmap,\, \convmapEstimateLow,\, \convmapEstimateHigh)$ have been defined in \eqref{eq:miscoveragerate}. We seek the smallest value $\calibparam$ satisfying $\riskfunCalibparam \leq \alpha$ with some confidence guarantees.
For this purpose, we consider a calibration set $\calibrationsetMM$ as introduced in Sect.~\ref{subsec:dfuq_cmm}. For a given error level $\delta \in \zeroexcloneexcl$, we compute Hoeffding's upper-confidence bound of $\riskfunCalibparam$, defined by
\begin{equation}
    \hoeffding := \frac1{\sizeCalibrationset} \sum_{i=1}^{\sizeCalibrationset} L\left(
        \convmap_i,\, \convmapEstimateLow_{i,\, \calibparam},\, \convmapEstimateHigh_{i,\, \calibparam}
    \right) + \sqrt{\frac{-\log \delta}{2\sizeCalibrationset}},
\label{eq:hoeffding}
\end{equation}
where $\convmapEstimateLow_{i,\, \calibparam}$ and $\convmapEstimateHigh_{i,\, \calibparam}$ satisfy \eqref{eq:lowerUpperBoundsCalib} with $\convmapEstimate := \convmapEstimate_i$ and $\residualEstimate := \residualEstimate_i$. These estimates are obtained from $\shearmap_i$ with the same method used to compute $\convmapEstimate$ and $\residualEstimate$. The RCPS approach is based on the following property, derived from Hoeffding's inequality:
\begin{equation}
    \Proba\left\{
        \hoeffdingRand < \riskfunCalibparam
    \right\} \leq \delta.
\label{eq:hoeffdingsineq}
\end{equation}
It controls the risk of underestimating the expected miscoverage rate by computing an empirical estimate on the calibration set.

We consider the following calibration parameter, computed on the calibration set:
\begin{equation}
    \calibparamRCPS{\alpha}{\delta} := \inf\set{\calibparam \in \mathR}{\hoeffding < \alpha},
\label{eq:rclambda}
\end{equation}
and the corresponding random vector $\calibparamRCPSRand{\alpha}{\delta}$.
By construction, the functions $\lambda \mapsto \riskfunCalibparam$ and $\lambda \mapsto \bighoeffding$
are monotone and non-increasing. Furthermore, we assume that the set $\set{\calibparam \in \mathR}{\riskfunCalibparam \leq \alpha}$ is non-empty. Then, the following result has been proven by \citet{Bates2021}, using \eqref{eq:hoeffdingsineq}:
\begin{equation}
    \Proba\,\Bigl\{
        \riskfunCalibparamRCPSRand > \alpha
    \Bigr\} \leq \delta.
\label{eq:riskcontrol0}
\end{equation}
Now, we consider the calibrated bounds with respect to $\calibparamRCPS{\alpha}{\delta}$:
\begin{equation}
    \convmapEstimateLowRCPS := \convmapEstimateLowCalibparamRCPS \qqand \convmapEstimateHighRCPS := \convmapEstimateHighCalibparamRCPS,
\end{equation}
where $\convmapEstimateLowCalibparamRCPS$ and $\convmapEstimateHighCalibparamRCPS$ satisfy \eqref{eq:lowerUpperBoundsCalib}. The corresponding random vectors are denoted by $\convmapEstimateLowRCPSRand$ and $\convmapEstimateHighRCPSRand$, respectively.
The random variable $\calibparamRCPSRand{\alpha}{\delta}$ only depends on the calibration set $\calibrationsetMMRand$, and therefore is independent of the miscoverage rate $L\bigl(
    \convmapRand,\, \convmapEstimateLowCalibparamRand,\, \convmapEstimateHighCalibparamRand
\bigr)$, for any $\calibparam \in \mathR$.
Consequently, the risk $\riskfunCalibparam$ defined in \eqref{eq:riskfun} can be expressed as follows:
\begin{align}
    \riskfunCalibparam
        &= \Bigcondexpval{
            L\left(
                \convmapRand,\, \convmapEstimateLowCalibparamRCPSRand,\, \convmapEstimateHighCalibparamRCPSRand
            \right)
        }{\calibparamRCPSRand{\alpha}{\delta} = \calibparam} \\
        &= \Bigcondexpval{
            L\left(
                \convmapRand,\, \convmapEstimateLowRCPSRand,\, \convmapEstimateHighRCPSRand
            \right)
        }{
            \calibparamRCPSRand{\alpha}{\delta} = \calibparam
        }.
\label{eq:riskfunCalib}
\end{align}
Finally, plugging \eqref{eq:riskfunCalib} into \eqref{eq:riskcontrol0} yields the following coverage guarantee for RCPS:
\begin{equation}
    \Proba\left\{
        \Bigcondexpval{
            L\left(
                \convmapRand,\, \convmapEstimateLowRCPSRand,\, \convmapEstimateHighRCPSRand
            \right)
        }{
            \calibparamRCPSRand{\alpha}{\delta}
        } > \alpha
    \right\} \leq \delta.
\label{eq:rcps_guarantee}
\end{equation}

\subsection{Differences between CQR and RCPS}
\label{subsec:dfuq_diff}

The theoretical coverage guarantees provided by the CQR and RCPS algorithms, respectively, \eqref{eq:cqr_guarantee1} and \eqref{eq:rcps_guarantee}, present some fundamental differences.
First, only CQR prevents overconservative solutions, due to the lower bound in \eqref{eq:cqr_guarantee1}. 
Moreover, in both cases, there exists a nonzero probability of accidentally selecting an out-of-distribution calibration set, leading to miscalibration and potentially large error rates. This possibility does not contradict the theoretical guarantees of either CQR or RCPS, because both approaches treat the calibration set as a set of random vectors. However, they handle this situation differently: the expected miscoverage rate is evaluated over the full distribution of calibration sets in the case of CQR, and conditionally on the calibration parameter $\calibparamRCPSRand{\alpha}{\delta}$---which depends on the calibration set $\calibrationsetMMRand$---in the case of RCPS. Consequently, unlike CQR, the risk of miscalibration due to selecting a too-small calibration parameter is explicit in \eqref{eq:rcps_guarantee}, and is controlled by the parameter $\delta$. Therefore, RCPS provides a higher degree of control over uncertainties. However, as confirmed by our experiments, it tends to produce more conservative prediction bounds than CQR, even for large values of $\delta$.

Finally, it is worth noticing that mass mapping with CQR, described in Sect.~\ref{subsec:dfuq_cmm}, requires $\imgsize^2$ calibration parameters gathered in the vector $\calibparamVecCQR{\alpha}$, which can be computed efficiently using linear algebra libraries such as NumPy and SciPy in Python. In contrast, the approach based on RCPS, described in Sect.~\ref{subsec:dfuq_rcps}, only requires one scalar calibration parameter $\calibparamRCPS{\alpha}{\delta}$.

\subsection{Differences with uncalibrated coverage guarantees}
\label{subsec:dfuq_diffuncalib}

By averaging over the set $\MM$ of active pixels, the target coverage guarantees before calibration---\eqref{eq:uq_freq} for the frequentist framework and \eqref{eq:uq_bayes} for the Bayesian framework---become, respectively,
\begin{equation}
    \Expval\left[
        L\left(
            \convmap,\, \convmapEstimateLowRand,\, \convmapEstimateHighRand
        \right)
    \right] \leq \alpha
\label{eq:uq_freq_imagewise}
\end{equation}
and
\begin{equation}
    \BigcondexpvalPrior{\mu}{
        L\left(
            \convmapRand,\, \convmapEstimateLowRand,\, \convmapEstimateHighRand
        \right)
    }{\shearmapRand = \shearmap} \leq \alpha,
\label{eq:uq_bayes_imagewise}
\end{equation}
where the conditional expected value $\bigcondexpvalPrior{\mu}{\cdot}{\shearmapRand = \shearmap}$ is defined with respect to the posterior density $\condpdf{\mu}{\convmap}{\shearmap}$ introduced in \eqref{eq:posteriordensity}.
In either \eqref{eq:cqr_guarantee1} and \eqref{eq:rcps_guarantee}, the expected miscoverage rate is evaluated over the full distribution of convergence maps, in contrast with \eqref{eq:uq_freq_imagewise} or \eqref{eq:uq_bayes_imagewise}, where it is taken conditionally on a single ground truth $\convmap$ (frequentist case) or observation $\shearmap$ (Bayesian case), respectively. Consequently, the calibration procedures do not prevent off-target miscoverage rates for specific images.


\section{Experiments}
\label{sec:exp}

\subsection{Experimental settings}
\label{subsec:exp_settings}

\subsubsection{Mass mapping methods}

In this paper, we tested CQR and RCPS on three mass mapping methods introduced in Sect.~\ref{subsec:sota_mm}---the KS inversion, the iterative Wiener algorithm, and MCALens. The uncalibrated uncertainty bounds $\convmapEstimateLow$ and $\convmapEstimateHigh$ were computed using the frequentist framework presented in Sect.~\ref{subsubsec:sota_uq_freq}. Specifically, for the KS method, the bounds were obtained analytically using \eqref{eq:2dconv}. In contrast, for the Wiener and MCALens methods, the bounds were computed using the Monte-Carlo method, by propagating $25$ noise realizations through the pipeline for each input shear map. The Python implementation of these algorithms is available on the CosmoStat GitHub repository.\footnote{\url{https://github.com/CosmoStat/cosmostat}}

\subsubsection{Simulated convergence maps}

In order to evaluate and compare the proposed methods, we used a calibration set $\calibrationsetMM$ of size $\sizeCalibrationset = 100$ and a test set $\testsetMM$ of size $\sizeCalibTestset - \sizeCalibrationset = 125$. The ground truths $\convmap_i$ were obtained using $\kappa$TNG cosmological hydrodynamic simulations \citep{Osato2021}, which include realizations of $5 \times 5 \deg^2$ convergence maps for various source redshifts at a $0.29$ arcmin per pixel resolution, assuming a flat $\Lambda$CDM Universe.

\begin{figure}
    \centering
    \includegraphics[width=0.8\columnwidth]{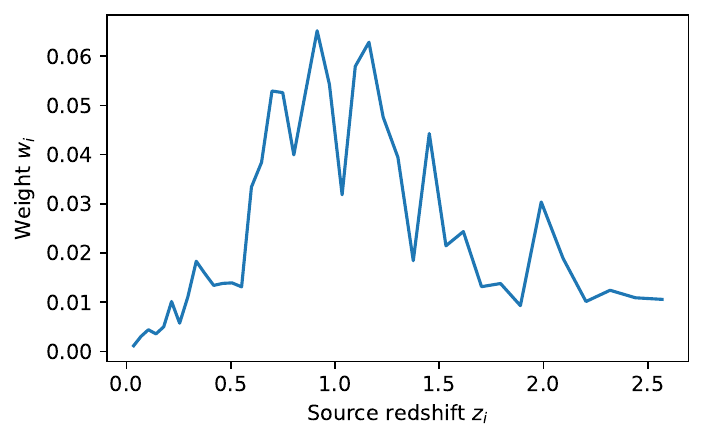}
    \caption{Distribution of redshifts used to combine simulated convergence maps from the $\kappa$TNG dataset.}
    \label{fig:z}
\end{figure}

For our experiments, we considered linear combinations of convergence maps according to a predefined redshift distribution, that we cropped to $306 \times 306$ pixels, with no overlap between them. This corresponds to an opening angle of $1.49 \deg$.
Formally, each convergence map $\convmap_i$ can be expressed as follows:%
\begin{equation}
    \convmap_i = \sum_{j=1}^{n_z} w_j \, \convmap_{ij},
\end{equation}
where $(\convmap_{ij})_{j=1}^{n_z}$ denotes a set of convergence map realizations at $n_z = 40$ source redshifts ranging from $3.45 \times 10^{-2}$ to $2.57$, and $w_j \in \zeroone$ denotes the weight assigned to the $j$-th source redshift $z_j$.
The weights $(w_j)_{j=1}^{n_z}$, which sum to $1$, were chosen to match the shape catalog used to compute the noisy shear maps $\shearmap_i$ (see Sect.~\ref{subsubsec:exp_settings_noise}). The weight distribution is displayed in Fig.~\ref{fig:z}.

\subsubsection{Noisy shear maps}
\label{subsubsec:exp_settings_noise}

\begin{figure}
    \centering
    \includegraphics[height=0.2\textheight]{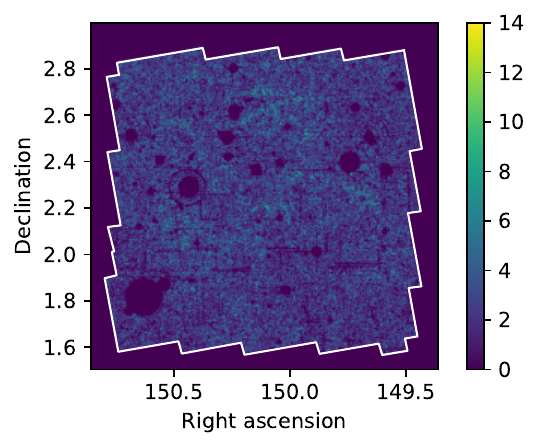}
    \caption{Number of galaxies per pixel using the S10 weak lensing shear catalog \citep{Schrabback2010}. All redshifts have been considered. The white borders delimitates the survey boundaries. Even within boundaries, some data are missing due to survey measurement masks.}
    \label{fig:ngal}
\end{figure}

\begin{figure*}
    \centering
    \begin{subfigure}[b]{0.32\textwidth}
        \centering
        \includegraphics[height=0.18\textheight]{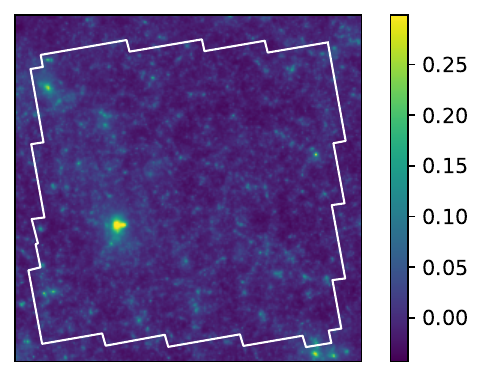}
        \caption{$\convmap_i$}
        \label{subfig:convshear_conv}
    \end{subfigure}
    \begin{subfigure}[b]{0.32\textwidth}
        \centering
        \includegraphics[height=0.18\textheight]{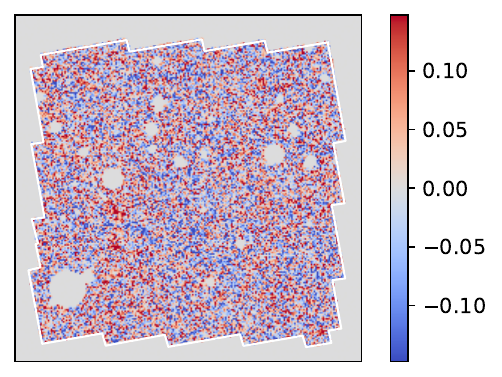}
        \caption{$\Real(\shearmap_i)$}
        \label{subfig:convshear_shear_re}
    \end{subfigure}
    \begin{subfigure}[b]{0.32\textwidth}
        \centering
        \includegraphics[height=0.18\textheight]{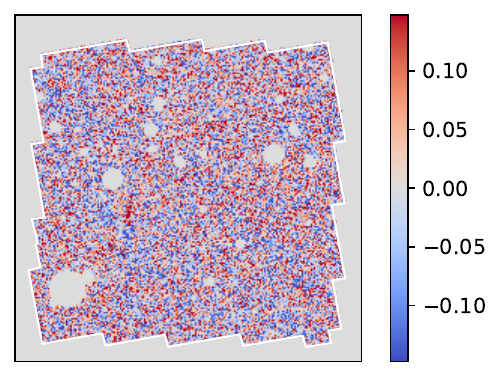}
        \caption{$\Imag(\shearmap_i)$}
        \label{subfig:convshear_shear_im}
    \end{subfigure}
    \caption{Example of simulated convergence map (Fig.~\ref{subfig:convshear_conv}), and the corresponding noisy shear map (Figs.~\ref{subfig:convshear_shear_re} and \ref{subfig:convshear_shear_im}).
    }
    \label{fig:convshear}
\end{figure*}

In accordance with \eqref{eq:invprob}, to generate noisy shear maps $\shearmap_i$ from $\convmap_i$, we applied the inverse Kaiser-Squires filter $\convToShear$ and added a zero-mean Gaussian noise $\noise_i$ with a diagonal covariance matrix $\covmatrNoise$ shared among all calibration and test input images. Its diagonal values were obtained by binning the weak lensing shear catalog created by \citet{Schrabback2010} \citep[referred to as ``S10'', following][]{Remy2023} at the $\kappa$TNG resolution ($0.29$ arcmin per pixel), and then applying \eqref{eq:covmatrnoise}. The intrinsic standard deviation $\stdval_{\ellip}$ was estimated from the measured ellipticities, and set to $0.28$. This is a reasonable estimation, assuming that the shear dispersion is negligible with respect to the dispersion of intrinsic ellipticities. The corresponding number of galaxies per pixel is represented in Fig.~\ref{fig:ngal}.

The S10 catalog was based on observations conducted by the NASA/ESA Hubble Space Telescope targeting the COSMOS field \citep{Scoville2007}, as well as photometric redshift measurements from \citet{Mobasher2007}. It contains an average of $32$ galaxies per square arcminute over a wide range of redshifts, which is consistent with what is expected from forthcoming surveys like Euclid. This number excludes the galaxies with redshifts below $3.45 \times 10^{-2}$ and above $2.57$, for the sake of consistency with the $\kappa$TNG dataset. It also ignores the inactive pixels $k \notin \MM$, without any observed galaxy. In these areas, we have simply set the shear map values to $0$, as explained in Appendix~\ref{subsec:appendix_pb_mask}. Consequently, we have omitted these regions from our statistical analyses.

An example of simulated convergence map $\convmap_i$ and corresponding noisy shear map $\shearmap_i$ is provided in Fig.~\ref{fig:convshear}.

\subsubsection{Reconstruction parameters}

The Wiener estimate $\convmapEstimate_{\wiener}$ as well as the Gaussian component $\convmapEstimate_{\gauss}$ of MCALens have been computed with a power spectrum (diagonal elements of $\fouriercovmatrConv$) empirically estimated from a dataset of $180$ simulated convergence maps distinct from the calibration and test sets. Moreover, the sparse component $\convmapEstimate_{\sparse}$ of MCALens was estimated using a starlet dictionary \citep{Starck2007}, which is well-suited for isotropic objects. The detection threshold for selecting the set of active wavelet coefficients was set to $4\sigma$. This threshold was selected based on visual evaluation, as it achieved a good balance between effectively detecting peaks and minimizing false positives.
For both iterative methods, the number of iterations was set to the default value of $12$. This choice was motivated by seeking a tradeoff between reconstruction accuracy and computational cost. Finally, the Kaiser-Squires maps were smoothed with a Gaussian filter $\BBS$ with a full width at half maximum (FWHM) of $2.4$ arcmin, following \citet{Starck2021}. To test the effect of the low-pass filter on reconstruction accuracy and error bar size, we also used a filter with half the original FWHM. These are referred to as strong and weak smoothing, respectively.

\subsubsection{UQ parameters}

We selected a target confidence level at $2\sigma$, corresponding to $\alpha \approx 0.046$. We notice that higher confidence levels require larger calibration sets, due to the finite-sample correction: the CQR algorithm selects the $(1 - \alpha)(1 + 1/\sizeCalibrationset)$-th empirical quantile of the conformity scores, which, by definition, must remain below $100\%$. Consequently, a $2\sigma$-confidence requires at least $\sizeCalibrationset = 21$ calibration samples, versus $\sizeCalibrationset = 370$ at $3\sigma$ and $\sizeCalibrationset = 15\,787$ at $4\sigma$.
Moreover, regarding the approach based on RCPS, we tested three values for the error level: $\delta = 0.05$, $0.2$, and $0.5$.

We implemented both approaches with several families of calibration functions, which is a novel aspect of this work:
\begin{equation}
    \calibfun_\calibparam: r \mapsto \max(r + \calibparam,\, 0)
    \label{eq:calibfun_add}
\end{equation}
as used by \citet{Romano2019}, and
\begin{equation}
    \calibfun_\calibparam: r \mapsto \lambda r
    \label{eq:calibfun_mult}
\end{equation}
as used by \citet{Angelopoulos2022b}. The former increases (or decreases) the size of the confidence intervals by the same value $\calibparam$ regardless of the initial size $2r$, whereas the latter adjusts the size proportionally to its initial value. This choice may have an influence on the average size of the calibrated confidence intervals. To push the analysis further, we also tested calibration functions with intermediate behaviors for the CQR approach, in the form
\begin{equation}
    \calibfun_\calibparam: r \mapsto r + b F_{\chi^2(k)}(r/a)(\calibparam - 1),
    \label{eq:calibfun_chisq}
\end{equation}
where $F_{\chi^2(k)}$ denotes the cumulative distribution function of a chi-squared distribution with $k$ degrees of freedom, and $a$ and $b$ denote positive real numbers. In practice, we tested values of $a$ ranging from $0.004$ to $0.012$, and adjusted $b$ to the maximum value such that $\calibfun_\calibparam$ remains non-decreasing for any $\calibparam \geq 0$.

\begin{figure*}
    \centering
    \begin{subfigure}[b]{0.32\textwidth}
        \centering
        \includegraphics[width=\textwidth]{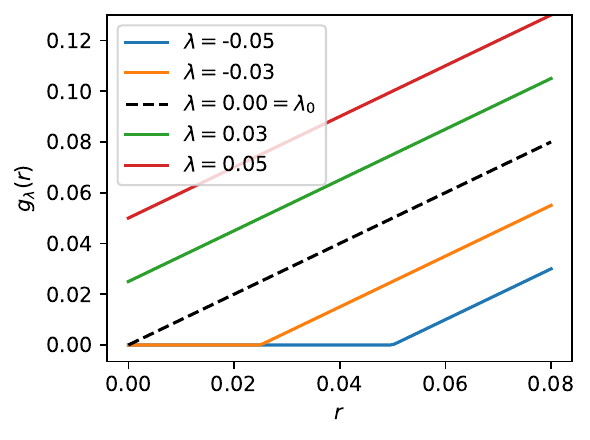}
        \caption{$\calibfun_\calibparam: r \mapsto \max(r + \calibparam,\, 0)$}
        \label{subfig:calibfun_1}
    \end{subfigure}
    \begin{subfigure}[b]{0.32\textwidth}
        \centering
        \includegraphics[width=\textwidth]{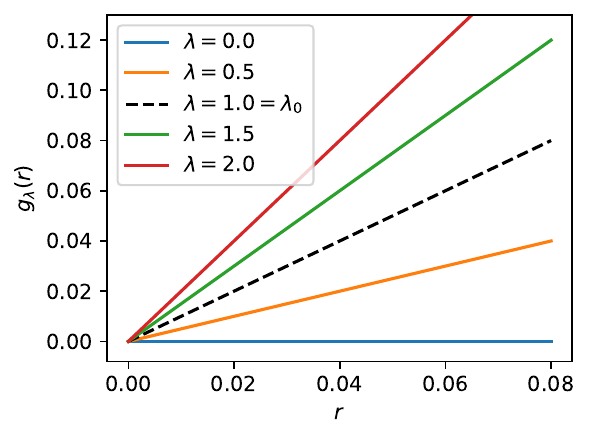}
        \caption{$\calibfun_\calibparam: r \mapsto \lambda r$}
        \label{subfig:calibfun_2}
    \end{subfigure}
    \begin{subfigure}[b]{0.32\textwidth}
        \centering
        \includegraphics[width=\textwidth]{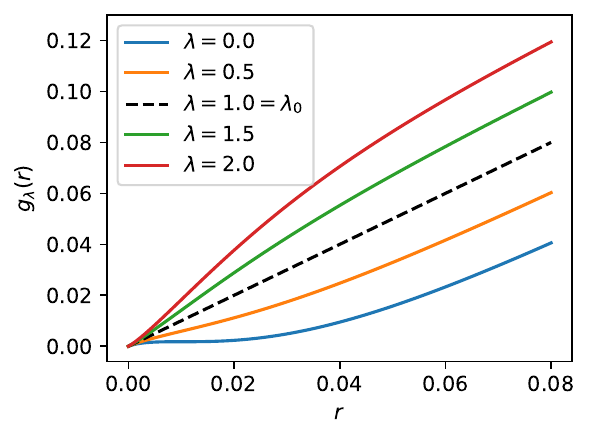}
        \caption{$\calibfun_\calibparam: r \mapsto r + b F_{\chi^2(k)}(r/a)(\calibparam - 1)$}
        \label{subfig:calibfun_3}
    \end{subfigure}
    \caption{Examples of families $(\calibfun_\calibparam)_\calibparam$ of calibration functions. In Fig.~\ref{subfig:calibfun_1} (additive calibration), $\calibparam$ ranges from $-\infty$ to $+\infty$ with $\calibparam_0 = 0$, whereas in Figs.~\ref{subfig:calibfun_2} (multiplicative calibration) and \ref{subfig:calibfun_3} (chi-squared calibration), $\calibparam$ ranges from $0$ to $+\infty$ with $\calibparam_0 = 1$. In Fig.~\ref{subfig:calibfun_3}, the number $k$ of degrees of freedom is set to $3$, the scaling factor $a$ is set to $0.01$, and the multiplicative factor $b$ is set to the maximum value such that $\calibfun_\calibparam$ remains non-decreasing for any $\calibparam \geq 0$, \ie, $b \approx 0.041$.}
    \label{fig:calibfun}
\end{figure*}

In the rest of the paper, the functions defined in \eqref{eq:calibfun_add}, \eqref{eq:calibfun_mult} and \eqref{eq:calibfun_chisq}, for which a visual representation is provided in Fig.~\ref{fig:calibfun}, are referred to as ``additive'', ``multiplicative'', and ``chi-squared'' calibration functions, respectively.

\subsection{Results}

\subsubsection{Visualization of reconstructed convergence maps}

A visual example of reconstructed convergence map, together with its prediction bounds before and after calibration, is provided in Fig.~\ref{fig:convestimates} (Appendix~\ref{sec:additional_fig_table}). A focus on the main peak-like structure is displayed in Fig.~\ref{fig:convestimates_highdensity} for the Wiener and MCALens estimates, after calibration with CQR.

We can observe that, for the KS and Wiener methods, the high-density region (bright spot in the convergence map) falls outside the confidence bounds, even after calibration (it is actually under-estimated). This is not in contradiction with the theoretical guarantees stated in \eqref{eq:cqr_guarantee} and \eqref{eq:rcps_guarantee}, because the expected miscoverage rate, computed across the whole set $\MM$ of active pixels, remains below the target $\alpha$, as explained in Sect.~\ref{subsubsec:exp_results_miscoveragerate}. However, these guarantees do not tell anything about the miscoverage rate of the higher-density regions specifically. A more detailed discussion on this topic is provided in Sect.~\ref{subsec:discussion_highdensity}.

In contrast, MCALens correctly predicts the high-density region. This observation is consistent with the fact that this algorithm was designed to accurately reconstruct the sparse component of the density field.

\subsubsection{Reconstruction accuracy}
\label{subsubsec:exp_results_mse}

In order to reproduce previously-established results, we measured, for each $i \in \range{\sizeCalibrationset + 1}{\sizeCalibTestset}$ in the test set, the root mean square error (RMSE) between the ground truth $\convmap_i$ and reconstructed convergence maps $\convmapEstimate_{\ks i}$, $\convmapEstimate_{\wiener i}$ and $\convmapEstimate_{\mcalens i}$, corresponding to the KS, Wiener and MCALens solutions, respectively. These metrics were computed on the set $\MM$ of active pixels, that is, within the survey boundaries, and for pixels with nonzero galaxies. We also measured the RMSE for high-density regions only, which are of greater importance when inferring cosmological parameters. More precisely, we only considered the pixels $k$ such that $|\convmap_i[k]| \geq 4.8 \times 10^{-2}$, which corresponds to a signal-to-noise (S/N) ratio above $0.25$ ($2.8\%$ of the total number of pixels). The results are displayed in Table~\ref{table:mse}.

\begin{table}
    \caption{Reconstruction accuracy}             
    \label{table:mse}
    \vspace{-5pt}
    \begin{centering}
    \small
    \begin{tabular}{ r | r r r r }
        \hline\hline
        & \multicolumn{4}{c}{RMSE ($\times 10^{-3}$)} \\
        \multicolumn{1}{c|}{Extent} & \multicolumn{1}{c}{KS1} & \multicolumn{1}{c}{KS2} & \multicolumn{1}{c}{Wiener} & \multicolumn{1}{c}{MCALens} \\
        \hline
        All pixels & $31.8 \pm 0.5$ & $21.1 \pm 0.8$ & $18.3 \pm 1.2$ & $\mathbf{18.0 \pm 1.0}$ \\
        High-density & $\mathbf{60.1 \pm 2.6}$ & $66.8 \pm 2.9$ & $72.3 \pm 3.0$ & $67.7 \pm 2.9$ \\
        \hline
    \end{tabular}
    \end{centering}
    \vspace{5pt}

    {\textbf{Note.} The RMSE is computed between each ground truth convergence map $\convmap_i$ and the corresponding reconstruction $\convmapEstimate_i$. The empirical means and standard deviations are computed over the test set $\testsetMM$ for each mass mapping method, and reported in this table. ``KS1'' and ``KS2'' correspond to the Kaiser-Squires estimators with weak and strong smoothing, respectively. The second row focuses on pixels $k$ such that $|\convmap_i[k]| \geq 4.8 \times 10^{-2}$.}
\end{table}

These results confirm that the Wiener and MCALens solutions achieve higher reconstruction accuracy compared to the KS solution, although the latter can be improved by increasing the standard deviation of the smoothing filter (see ``KS1'' versus ``KS2''). Additionally,
MCALens outperforms Wiener by $1.6\%$.
This difference significantly increases when focusing on high-density regions ($6.3\%$ improvement).

\subsubsection{Miscoverage rate}
\label{subsubsec:exp_results_miscoveragerate}

\begin{figure*}
    \centering
    \begin{subfigure}[t]{0.5\textwidth}
        \vtop{\null\hbox{\includegraphics[width=\textwidth]{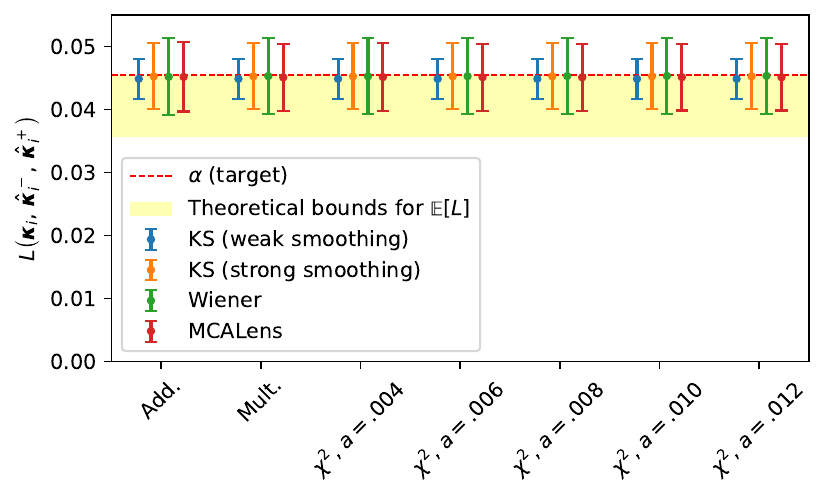}}}
        \caption{CQR}
        \label{subfig:errs_cqr}
    \end{subfigure}
    \begin{subfigure}[t]{0.49\textwidth}
        \vtop{\null\hbox{\includegraphics[width=\textwidth]{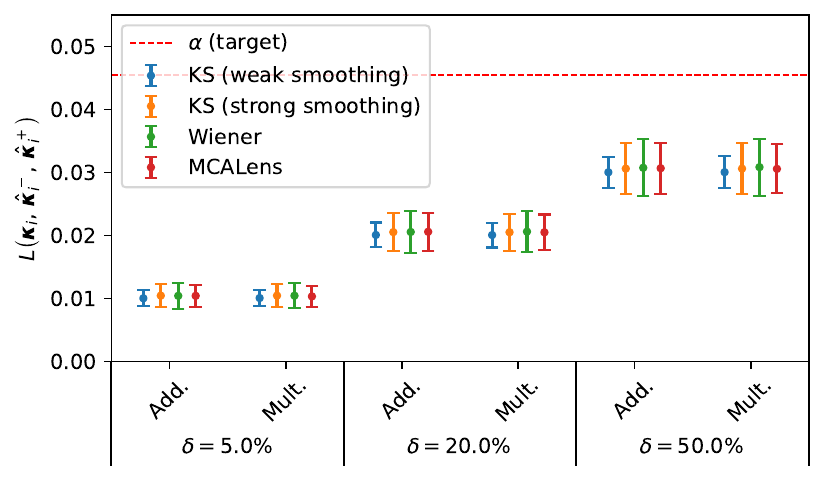}}}
        \caption{RCPS}
        \label{subfig:errs_rcps}
    \end{subfigure}
    \caption{Empirical miscoverage rate after calibration with CQR (Fig.~\ref{subfig:errs_cqr}) and RCPS (Fig.~\ref{subfig:errs_rcps}). The means and standard deviations are computed over the test set $\testsetMM$, for various error levels $\delta$ (RCPS only), and various families of calibration functions $(\calibfun_\calibparam)_\calibparam$. More details are provided in the note of Table~\ref{table:results} (Appendix~\ref{sec:additional_fig_table}). The theoretical bounds for CQR, introduced in \eqref{eq:cqr_guarantee1}, are represented by the yellow area in Fig.~\ref{subfig:errs_cqr}. These plots indicate that CQR achieves miscoverage rates that are, on average, close to the target $\alpha$, whereas RCPS tends to produce overconservative bounds, even for large values of $\delta$.}
    \label{fig:errs}
\end{figure*}

For each $i \in \range{\sizeCalibrationset + 1}{\sizeCalibTestset}$ in the test set, we have measured the empirical miscoverage rate $L(\convmap_i,\, \convmapEstimateLow_i,\, \convmapEstimateHigh_i)$ such as introduced in \eqref{eq:miscoveragerate}, where the (uncalibrated) lower- and upper-bounds $\convmapEstimateLow_i$ and $\convmapEstimateHigh_i$ have been computed following \eqref{eq:lowerUpperBounds}, using each of the three reconstruction methods. Then, after calibrating the bounds using CQR and RCPS on the calibration set $\calibrationsetMM$, we have computed $L\bigl(
    \convmap_i,\, \convmapEstimateLow_{\cqr i},\, \convmapEstimateHigh_{\cqr i}
\bigr)$ and $L\bigl(
    \convmap_i,\, \convmapEstimateLow_{\rc i},\, \convmapEstimateHigh_{\rc i}
\bigr)$ on the test set $\testsetMM$. Finally, we compared the results with the theoretical guarantees stated in \eqref{eq:cqr_guarantee1} and \eqref{eq:rcps_guarantee}, respectively. The results are reported in Table~\ref{table:results} (Appendix~\ref{sec:additional_fig_table}), and plotted in Fig.~\ref{fig:errs}.

In the case of CQR, the empirical mean of the miscoverage rate falls between the theoretical bounds (yellow area). This is in line with the theoretical guarantee stated in \eqref{eq:cqr_guarantee1}. We also notice that the lower bound seems overly conservative. It should be noted that the expected value from \eqref{eq:cqr_guarantee1} covers uncertainties over the convergence map $\convmapRand$, the noise $\noiseRand$, but also the calibration set $\calibrationsetMMRand$. In our experiments, we only considered one realization $\calibrationsetMM$ of the calibration set, which, as explained in Sect.~\ref{subsec:dfuq_diff}, may result in miscalibration and above-target error rates. This effect could be mitigated by increasing the size $\sizeCalibrationset$ of the calibration set, or with bootstrapping.

Unlike CQR, the RCPS approach controls the risk of picking a statistically-deviant calibration set, which could lead to undercoverage, by introducing an additional parameter $\delta$. However, even with large values of $\delta$, the calibrated bounds tend to be overly conservative. For example, with $\delta = 0.5$, one would expect the average miscoverage rate to fluctuate around the target $\alpha$ in approximately $50\%$ of experiments when repeating the protocol with different calibration sets. In practice, the computed miscoverage rates consistently fall well below the target. Contrary to CQR, RCPS does not prevent situations of overcoverage.

\subsubsection{Mean length of prediction intervals}
\label{subsubsec:exp_results_meanlength}

\begin{figure*}
    \centering
    \begin{subfigure}[t]{0.5\textwidth}
        \vtop{\null\hbox{\includegraphics[width=\textwidth]{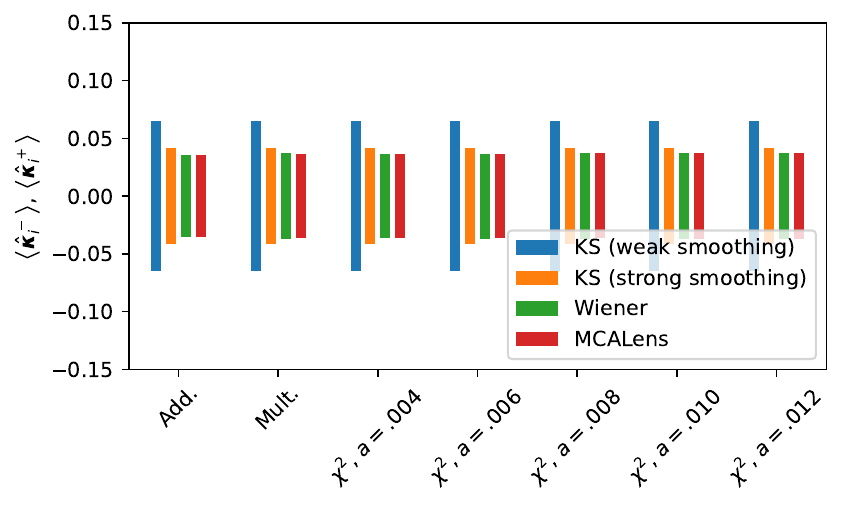}}}
        \caption{CQR}
        \label{subfig:confidencebounds_cqr}
    \end{subfigure}
    \begin{subfigure}[t]{0.49\textwidth}
        \vtop{\null\hbox{\includegraphics[width=\textwidth]{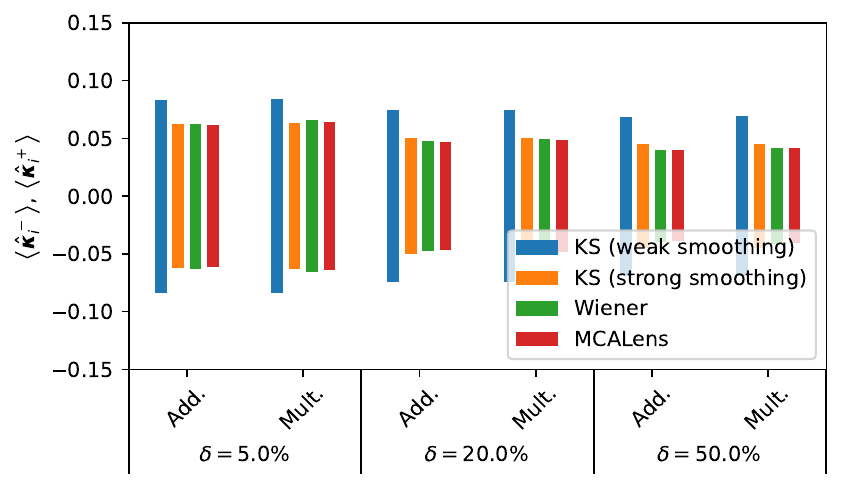}}}
        \caption{RCPS}
        \label{subfig:confidencebounds_rcps}
    \end{subfigure}
    \caption{Mean values for the lower and upper confidence bounds, computed over the span of each image. The bar sizes represent the mean length of the error bars. As in Fig.~\ref{fig:errs}, the values are computed over the test set $\testsetMM$, for various families of calibration functions $(\calibfun_\calibparam)_\calibparam$. After calibration with CQR, the choice of mass mapping method does not influence the average miscoverage rate (Fig.~\ref{subfig:errs_cqr}), but significantly affects the size of the error bars (Fig.~\ref{subfig:confidencebounds_cqr}). Additionally, RCPS, which tends to produce overconservative prediction bounds (Fig.~\ref{subfig:errs_rcps}), yields larger error bars than CQR (Fig.~\ref{subfig:confidencebounds_rcps} vs \ref{subfig:confidencebounds_cqr}).}
    \label{fig:confidencebounds}
\end{figure*}

The mean length of the prediction intervals has been computed over each image in the test set, before and after calibration, for each mass mapping method, each calibration method (CQR and RCPS), and each family of calibration functions. The results are reported in Table~\ref{table:results} (Appendix~\ref{sec:additional_fig_table}), and plotted in Fig.~\ref{fig:confidencebounds}.

We observe that the choice of mass mapping method influences the size of the confidence intervals. Specifically, the KS solution yields larger calibrated error bars than either the Wiener or MCALens solutions, particularly in the weak-smoothing scenario (``KS1''). Additionally, the smallest confidence intervals are obtained with MCALens, using CQR with the additive family of calibration functions (in bold in Table~\ref{table:results}). This family consistently produces equal or smaller confidence intervals compared to the multiplicative or chi-squared families, for both CQR and RCPS calibration procedures. The modest improvement of the MCALens solution compared to Wiener ($0.24\%$ reduction in the mean length) can be attributed to the overall similarity between MCALens and Wiener outputs, with MCALens demonstrating higher reconstruction accuracy only in a few peak-like structures, which are essential for inferring cosmological parameters. Consequently, averaging the results over all pixels tends to hide essential properties of MCALens. A more detailed discussion on high-density regions is provided in Sect.~\ref{subsec:discussion_highdensity}.

As explained above, RCPS produces overly conservative bounds, which has a strong impact on the size of the confidence intervals. Smaller confidence intervals could be obtained by increasing the size $\sizeCalibrationset$ of the calibration set, since the Hoeffding's upper-confidence bound \eqref{eq:hoeffding} decreases with increased values of $\sizeCalibrationset$. Therefore, CQR seems more appropriate if the computational resources or available data are limited.

\section{Discussion}
\label{sec:discussion}

\subsection{Distribution-free vs Bayesian uncertainties}

Extending distribution-free UQ to data-driven mass mapping methods (see Sect.~\ref{subsec:sota_mm}) is straightforward, provided one has access to a ``first guess'', an initial quantification of uncertainty. Such initial uncertainty bounds can be obtained following the Bayesian framework presented in Sect.~\ref{subsubsec:sota_uq_bayes}, which provides theoretical guarantees similar to \eqref{eq:uq_bayes_imagewise}.

However, as explained in Sect.~\ref{subsubsec:sota_uq_limits}, this heavily relies on a proper choice of the prior distribution $\mu$, and if the latter differs from the unknown oracle distribution $\mu^\ast$, the empirical mean of the miscoverage rate, computed over the test set, may significantly diverge from \eqref{eq:uq_bayes_imagewise}.
Applying a post-processing calibration step, such as CQR or RCPS, allows for obtaining the guarantees stated in \eqref{eq:cqr_guarantee1} or \eqref{eq:rcps_guarantee}, respectively. In these expressions, the expected value assumes $\convmapRand \sim \mu^\ast$, even though the underlying distribution $\mu^\ast$ remains implicit---hence the term ``distribution-free''. The only sufficient conditions for these properties to be satisfied are that the data from the calibration and test sets $\datasetMMRand{1}{\sizeCalibTestset}$ are exchangeable, and that the conformity scores are almost surely distinct.

\subsection{Beyond pixelwise uncertainty}
\label{subsec:discussion_beyondpixelwise}

This paper focuses on estimating error bars per pixel. While this approach is easily interpretable, it lacks some essential properties. First, it is important to consider the purpose of mass mapping and the type of information required. For instance, to infer cosmological parameters, several studies have proposed using peak counts \citep{Kratochvil2010,Marian2013,Shan2014,Liu2015}, convolutional neural networks
that exploit information in the gradient around peaks \citep{Ribli2019a,Ribli2019}, or neural compression of peak summary statistics \citep{Jeffrey2021}. Consequently, regions of higher density are of particular interest and should be treated with special attention.

In this context, alternative UQ schemes could be considered. Among these are the Bayesian hypothesis testing of structure \citep{Cai2018a,Cai2018,Repetti2019,Price2021}, which can assess whether a given structure in an image is a reconstruction artifact or has some physical significance. Additionally, a computationally efficient approach, based on wavelet decomposition and thresholding of the MAP reconstruction, has been recently proposed by \citet{Liaudat2023}. This method estimates errors at various scales, highlighting the different structures of the reconstructed image.
In fact, these ideas can be traced back to earlier work in the stochastic geometry literature \citep[see, \eg, the book by][]{Adler2007a}, with applications in neuroimaging \citep[\eg,][]{Fadili2004}. Another potential direction is the application of calibration procedures to blob detection \citep{Lindeberg1993}, for which UQ methods have been recently developed by \citet{Parzer2023,Parzer2024}.

Another drawback of pixelwise UQ is that it does not account for correlations between pixels. For instance, whether a given pixel has been accurately predicted can impact the uncertainty bounds of the neighboring pixels, a property not reflected in per-pixel error bars. This concern intersects with the previous one, as it raises the question of correctly identifying the structures of interest in the reconstructed image, which typically span several pixels.
To address this issue, \citet{Belhasin2024} proposed applying UQ after decomposing the reconstructed images through principal component analysis. They also applied RCPS in this context, demonstrating that calibration methods can be used beyond the per-pixel framework.

\subsection{Focus on higher-density regions}
\label{subsec:discussion_highdensity}

\begin{figure}
    \centering
    \includegraphics[width=\columnwidth]{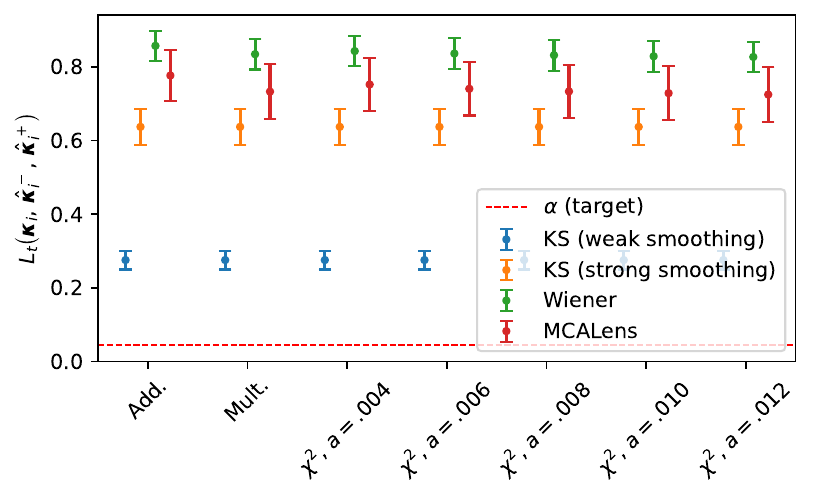}
    \caption{Empirical miscoverage rate after calibration with CQR, computed on regions of higher density. Following Sect.~\ref{subsubsec:exp_results_mse}, only pixels $k$ with $|\convmap_i[k]| \geq t = 4.8 \times 10^{-2}$ were considered. Unlike in Fig.~\ref{subfig:errs_cqr}, where the miscoverage rate is computed over all pixels, the measured values in this case deviate drastically from the target. Consequently, the size of the error bars is underestimated in these areas.}
    \label{fig:errs_highdensity}
\end{figure}

\begin{figure}
    \centering
    \begin{subfigure}{\columnwidth}
        \centering
        \includegraphics[width=0.4\textwidth]{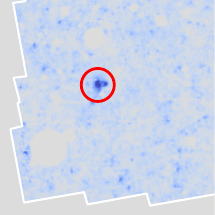}
        \hspace{10pt}
        \includegraphics[width=0.4\textwidth]{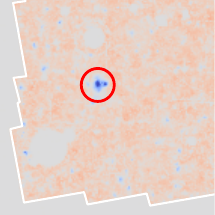}
        \caption{Wiener}
        \vspace{5pt}
        \label{subfig:convestimates_highdensity_wiener}
    \end{subfigure}
    \begin{subfigure}{\columnwidth}
        \centering
        \includegraphics[width=0.4\textwidth]{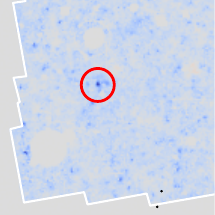}
        \hspace{10pt}
        \includegraphics[width=0.4\textwidth]{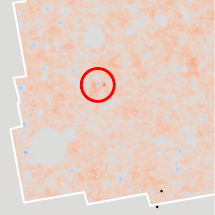}
        \caption{MCALens}
        \label{subfig:convestimates_highdensity_mcalens}
    \end{subfigure}
    \caption{Lower and upper bounds after calibration with CQR, with a focus on a peak-like structure. See Fig.~\ref{fig:convestimates} for more details. In the Wiener solution, the ground truth $\convmap[k]$ remains above $\convmapEstimateHigh[k]$ (miscoverage), whereas MCALens provides better coverage. This observation aligns with Fig.~\ref{fig:errs_highdensity}.}
    \label{fig:convestimates_highdensity}
\end{figure}

As confirmed by our experiments, CQR provides guarantees on the miscoverage rate $L(\convmap,\, \convmapEstimateLow,\, \convmapEstimateHigh)$ defined in \eqref{eq:miscoveragerate}. However, this score is computed by considering all active pixels $k \in \MM$. This may hide disparities caused by latent factors such as the local density of the convergence field. To support this claim, we computed the miscoverage rate filtered on regions of higher density. More precisely, for any $i \in \range{\sizeCalibrationset + 1}{\sizeCalibTestset}$, we only considered pixels $k$ such that $|\convmap_i[k]| \geq t$ for a given threshold $t \in \mathR$, set to $4.8 \times 10^{-2}$ in our experiments (S/N ratio above $0.25$), similarly to Sect.~\ref{subsubsec:exp_results_mse}. The corresponding metric is defined by%
\begin{equation}
    L_{t}(\convmap,\, \convmapEstimateLow,\, \convmapEstimateHigh) :=
    \frac{
        \card\,\Bigset{
            k \in \MM_t(\convmap)
        }{
            \convmap[k] \notin \bigl[
                \convmapEstimateLow[k],\, \convmapEstimateHigh[k]
            \bigr]
        }
    }{\card \MM_t(\convmap)},
\label{eq:miscoveragerate_filtered}
\end{equation}
with
\begin{equation}
    \MM_t(\convmap) := \Bigset{k \in \MM}{\convmap[k] \geq t}.
\end{equation}
The results, displayed in Fig.~\ref{fig:errs_highdensity}, indicate that the filtered miscoverage rate remains well above the target $\alpha$, even after calibration with CQR. This phenomenon can also be visualized in Fig.~\ref{fig:convestimates_highdensity} for the Wiener method: the bright area remains underestimated.

As highlighted in Sect.~\ref{subsec:discussion_beyondpixelwise}, these high-density regions are important for inferring cosmological parameters. Therefore, the method could be improved by incorporating this additional constraint. An interesting direction involves conformal prediction masks, as introduced by \citet{Kutiel2023}. The method consists of masking the regions of low uncertainty in a given image, allowing a focus on the regions of higher uncertainty---corresponding to high-density regions in the case of weak lensing mass mapping.

\begin{figure}
    \centering
    \includegraphics[trim=0 0 0 7pt, clip, width=\columnwidth]{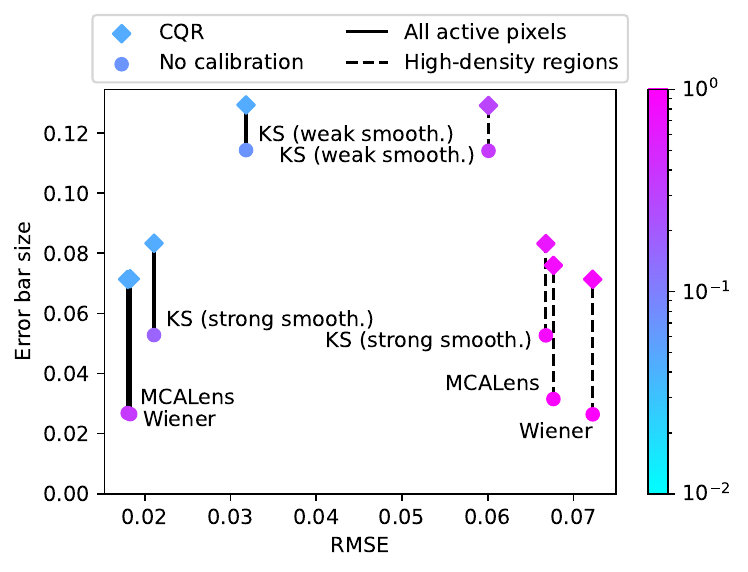}
    \caption{Mean length of prediction intervals (Sect.~\ref{subsubsec:exp_results_meanlength}) plotted against RMSE (Sect.~\ref{subsubsec:exp_results_mse}), both before and after calibration using additive CQR, with and without filtering on high-density regions. Marker colors indicate empirical miscoverage rates (Sect.~\ref{subsubsec:exp_results_miscoveragerate}). The coverage guarantee \eqref{eq:cqr_guarantee1} applies only after calibration (blue diamonds atop the solid lines). We notice that, when filtering on high-density regions, the coverage property no longer holds (pink diamonds atop the dashed lines).}
    \label{fig:summary}
\end{figure}

Fig.~\ref{fig:summary} presents a scatter plot summarizing the various metrics discussed in this study.
It indicates that Wiener and MCALens show similar error bar size and RMSE when computed across the entire set $\MM$ of active pixels. From these metrics' perspective, both methods outperform the KS estimators. However, when focusing on high-density regions, the Wiener solution exhibits poorer performance in terms of reconstruction accuracy (RMSE) and miscoverage rate, compared to MCALens and KS. As a result, MCALens achieves a balance between overall and high-density-specific performance. Additionally, the choice of smoothing filter significantly influences the reconstruction accuracy and error bar size of the KS estimator.




\section{Conclusion}
\label{sec:concl}

In this work, we emphasized the need for adjusting uncertainty estimates for reconstructed convergence maps, whether computed through frequentist or Bayesian frameworks.
To address this, we built on two recent distribution-free calibration methods---conformalized quantile regression (CQR) and risk-controlling prediction sets (RCPS)---to obtain error bars with valid coverage guarantees.

This paper presents several key innovations. First, while RCPS has been applied to inverse problems in other contexts, CQR was originally designed for scalar regression and required adaptation for this specific application. Second, no prior work has compared the two methods directly. Finally, we investigated various families of calibration functions for both approaches.

Our experiments led to three key findings. First, RCPS tends to produce overconservative confidence intervals, whereas CQR allows for more accurate---and smaller---error bars. Second, the choice of mass mapping method significantly influences the size of these error bars: a $45\%$ reduction for MCALens compared to KS with weak smoothing, a $14\%$ reduction compared to KS with strong smoothing, and a $0.24\%$ reduction compared to Wiener. Finally, this choice also affects the reconstruction accuracy, especially around peak-like structures, where MCALens outperforms Wiener by $6.3\%$.

We emphasize that these calibration approaches are not limited to the model-driven mass mapping methods discussed in this paper; they are applicable to any mass mapping method, including those based on deep learning.

To conclude and guide future research, we acknowledge several limitations of this study. First, particular attention must be given to the intensity of peak-like structures, which are essential for inferring cosmological parameters. As it stands, neither of the two calibration methods effectively prevents miscoverage in these high-density regions.
Additionally, this work could be expanded to include a broader range of uncertainty estimates beyond per-pixel error bars. Finally, the calibration and resulting uncertainty bounds were derived for a specific set of cosmological parameters. While we postulate that variations in cosmology have only a marginal impact on the results, this hypothesis needs further investigation.

\begin{acknowledgements}
    This work was funded by the TITAN ERA Chair project (contract no.\@ 101086741) within the Horizon Europe Framework Program of the European Commission, and the  Agence Nationale de la Recherche (ANR-22-CE31-0014-01 TOSCA and ANR-18-CE31-0009 SPHERES).
\end{acknowledgements}

\section*{Data availability}

To ensure reproducibility, all software, scripts, and notebooks used in this study are available on GitHub.\footnote{\url{https://github.com/hubert-leterme/weaklensing_uq.git}}

%
\bibliographystyle{bibtex/aa} 
\bibliography{bibtex/refs} 
%

\begin{appendix}

\section{Mass mapping: problem formulation}
\label{sec:appendix_pb}

The following formalism has been outlined by \citet{Kilbinger2015}.

\subsection{Weak lensing and mass mapping}

Consider a light ray emitted by an extended source (typically, a galaxy), observed at coordinates $\btheta \in \mathR^2$ in the sky. In the presence of (inhomogeneous) matter density that we seek to estimate, the light ray undergoes continuous deflection as it travels through space, a phenomenon known as gravitational lensing. In the absence of lensing, it would be seen by the observer at coordinates $\bbeta(\btheta)$. Assuming deflections are small enough (weak lensing regime), one can write
\begin{equation}
    \bbeta(\btheta) = \btheta - \nabla \psi(\btheta),
\label{eq:coordtransf}
\end{equation}
where $\psi: \mathR^2 \to \mathR$ denotes a lensing potential. Now, consider another light ray emitted by the same galaxy, observed at coordinates $\btheta + \delta\btheta$. A first-order approximation yields
\begin{equation}
    \bbeta(\btheta + \delta\btheta) - \bbeta(\btheta) = \BBJ(\btheta) \cdot \delta\btheta,
\end{equation}
where the Jacobian $\BBJ: \mathR^2 \to \mathR^{2 \times 2}$ (also referred to as amplification matrix) satisfies
\begin{equation}
    \BBJ := \begin{pmatrix}
        1 - \kappa - \gamma_1 & -\gamma_2 \\
        -\gamma_2 & 1 - \kappa + \gamma_1
    \end{pmatrix} = (1 - \kappa) \begin{pmatrix}
        1 - \reducedshear_1 & -\reducedshear_2 \\
        -\reducedshear_2 & 1 - \reducedshear_1
    \end{pmatrix},
\label{eq:jacobian}
\end{equation}
where we have introduced the fields $\kappa,\, \gamma_1,\ \gamma_2: \mathR^2 \to \mathR$ such that
\begin{equation}
    \begin{matrix}
        \kappa := \frac12 (\partial_1^2 \psi + \partial_2^2 \psi); & \gamma_1 := \frac12 (\partial_1^2 \psi - \partial_2^2 \psi); & \gamma_2 := \partial_1\partial_2\psi,
    \end{matrix}
    \label{eq:convergence_shear}
\end{equation}
and $\reducedshear_i := \gamma_i / (1 + \kappa)$, $i = 1,\, 2$.
For any $\btheta \in \mathR^2$, $\kappa(\btheta) \in \mathR$, referred to as the convergence, translates into an isotropic dilation of the source, and $\gamma(\btheta) := \gamma_1(\btheta) + i\gamma_2(\btheta) \in \mathC$, referred to as the shear, causes anisotropic stretching of the image, typically transforming a circle into an ellipse. Finally, $\reducedshear(\btheta) := \reducedshear_1(\btheta) + i\reducedshear_2(\btheta)$ is referred to as the reduced shear.

The goal is now to estimate the projected mass between the observer and the observed galaxies in any direction.
In the weak lensing regime, it can be shown that the convergence $\kappa(\btheta)$ is roughly proportional to the (weighted) matter density projected along the line of sight between the observer and the source. Therefore, mass mapping can be performed by estimating the convergence map $\kappa: \mathR^2 \to \mathR$ from the observation of galaxies at a given redshift.

\subsection{Relationship between shear and convergence}

Although not directly measurable, $\kappa$ can be retrieved from the shear $\gamma$, up to an additive constant. In the flat sky limit, solving the PDEs \eqref{eq:convergence_shear} yields the following relationship in the Fourier space:
\begin{equation}
    \forall \bnu \neq \bzero,\, \tilde\gamma(\bnu) = \frac1{\pi} \tilde{\mathcal D}(\bnu) \, \tilde \kappa(\bnu), \;\mbox{with}\; \tilde{\mathcal D}: \bnu \mapsto \pi \frac{(\nu_1 + i \nu_2)^2}{\nu_1^2 + \nu_2^2}.
\label{eq:convprod_fourier}
\end{equation}
Denoting by $\mathcal F$ the Fourier transform operator, this yields
\begin{equation}
    \gamma(\btheta) = \frac1{\pi} \mathcal F^{-1} \! \left(
        \tilde{\mathcal D} \tilde \kappa
    \right)\!(\btheta).
\label{eq:invfouriertransf}
\end{equation}

Now, we discretize \eqref{eq:invfouriertransf} on a square grid of size $\imgsize^2 \in \mathN$, and adopt a vector-matrix notation in which the shear and convergence maps $\shearmap \in \mathC^{\imgsize^2}$ and $\convmap \in \mathR^{\imgsize^2}$ are flattened and thus represented as one-dimensional vectors. By neglecting discretization and boundary effects, we get the following relation:
\begin{equation}
    \shearmap = \convToShear \convmap + \gamma_0, \qqwith \convToShear := \fouriermatrHerm \BBP \fouriermatr,
\label{eq:invprob_nonoise}
\end{equation}
where $\fouriermatr$ and its Hermitian transpose $\fouriermatrHerm \in \mathC^{\imgsize^2 \times \imgsize^2}$ respectively encode the discrete Fourier and inverse Fourier transforms, and $\BBP \in \mathC^{\imgsize^2 \times \imgsize^2}$ is a diagonal matrix obtained by discretizing $\tilde{\mathcal D} / \pi$ on the same grid: for any $(k_1,\, k_2) \neq (0,\, 0)$,
\begin{equation}
    \BBP\left[
        \imgsize^2 k_1 + k_2,\, \imgsize^2 k_1 + k_2
    \right] := \frac{(k_1 + i k_2)^2}{k_1^2 + k_2^2}.
\end{equation}
The constant $\gamma_0 := \langle \shearmap \rangle \in \mathC$ (mean value of the shear map) accounts for the fact that $\tilde{\mathcal D}$ is discontinuous in $\bzero$, thereby leaving $\BBP[0,\, 0]$ undefined when discretizing (mass-sheet degeneracy). In practice, we therefore set $\BBP[0,\, 0] = 0$ and add this constant to the equation.

Due to mass-sheet degeneracy, the operator $\convToShear$ is not invertible. Therefore, assuming the shear map $\shearmap$ is known, the convergence map $\convmap$ can only be retrieved up to an additive constant:
\begin{equation}
    \convmap = \convToShearPseudoinv \shearmap + \kappa_0, \qqwith \convToShearPseudoinv := \fouriermatrHerm \BBP^\dagger \fouriermatr,
\label{eq:shear2conv}
\end{equation}
where $\kappa_0 := \langle \convmap \rangle \in \mathR$ denotes the mean value of the convergence map, and $\BBP^\dagger \in \mathC^{\imgsize^2 \times \imgsize^2}$ denotes a diagonal matrix defined by $\BBP^\dagger[0,\, 0] := 0$ and, for all $k \neq 0$, by $\BBP^\dagger[k, k] := \BBP[k, k]^{-1}$.
In practice, we will therefore estimate the variations of $\convmap$ around its mean value.

\subsection{Shear measurement}

The intrinsic ellipticity of a galaxy is characterized by a complex number $\epsilon^s \in \mathC$. In the weak lensing regime, the observed ellipticity $\epsilon$ approximately satisfies
\begin{equation}
    \epsilon = \reducedshear + \epsilon^s,
\label{eq:ellipticity}
\end{equation}
where, as a reminder, $\reducedshear \in \mathC$ denotes the reduced shear, which varies with the galaxy coordinates $\btheta$. A common approximation in weak lensing is to consider $\kappa \ll 1$, and therefore $\reducedshear \approx \gamma$. In the remaining of the section, we will replace $\reducedshear$ by the unreduced shear $\gamma$.
Assuming the intrinsic galaxy ellipticity has no preferred orientation, the expected value of $\epsilon^s$ vanishes, which yields $\Expval[\epsilon] = \Expval[\gamma]$.
Therefore, the observed ellipticity $\epsilon$ is an unbiased estimator of $\gamma$.

In practice, we consider a portion of the sky on which one seek to perform mass mapping, which we subdivide into $\imgsize^2$ small regions (or bins), each of which associated to a pixel in the discretized shear map $\shearmap$ and convergence map $\convmap$. For each bin $k$, we measure the ellipticity of $\ngalperpix{k}
$ galaxies at a given redshift. Then, \eqref{eq:ellipticity} yields
\begin{equation}
    \frac1{\ngalperpix{k}} \sum_{i=1}^{\ngalperpix{k}} \epsilon_{k,\, i} = \frac1{\ngalperpix{k}} \sum_{i=1}^{\ngalperpix{k}} \gamma_{k,\, i} + \frac1{\ngalperpix{k}} \sum_{i=1}^{\ngalperpix{k}} \epsilon^s_{k,\, i},
\label{eq:meanellipticity}
\end{equation}
where $\gamma_{k,\, i}$ denotes the shear, and $\epsilon_{k,\, i}$ and $\epsilon^s_{k,\, i}$ respectively denote the intrinsic and measured ellipticity of the $i$-th galaxy within the $k$-th bin. Assuming that the bins are small enough to maintain an approximate consistancy of the shear within each, we get
\begin{equation}
    \frac1{\ngalperpix{k}} \sum_{i=1}^{\ngalperpix{k}} \gamma_{k,\, i} = \shearmap[k].
\label{eq:meanshear}
\end{equation}
Moreover, we introduce
\begin{equation}
    \ellmap[k] := \frac1{\ngalperpix{k}} \sum_{i=1}^{\ngalperpix{k}} \epsilon_{k,\, i}
        \qand
    \noise[k] := \frac1{\ngalperpix{k}} \sum_{i=1}^{\ngalperpix{k}} \epsilon^s_{k,\, i},
\label{eq:shear_noise}
\end{equation}
in which $\ellmap$ is an unbiased estimator of $\shearmap$, and $\noise$ is the realization of a Gaussian noise with zero mean and diagonal covariance matrix $\covmatrNoise \in \mathC^{\imgsize^2 \times \imgsize^2}$, satisfying
\begin{equation}
    \covmatrNoise[k,\, k] = \stdval_{\ellip} / \sqrt{\ngalperpix{k}},
\label{eq:covmatrnoise}
\end{equation}
where $\stdval_{\ellip} \in \mathC$ denotes the standard deviation of the intrinsic ellipticities. In practice, this quantity is estimated by measuring the standard deviation of the observed ellipticities across the sky. In practice, the number of galaxies per bin may be quite small ($N_k \sim 10$), which yields a very low S/N ratio.

Then, combining \eqref{eq:invprob_nonoise} and \eqref{eq:meanellipticity}-\eqref{eq:shear_noise} yields
\begin{equation}
    \ellmap = \convToShear \convmap + \noise + \gamma_0.
\end{equation}
Since the noise $\noise$ has zero-mean, we have $\gamma_0 \approx \langle \ellmap \rangle$. Therefore, by replacing the true unknown shear map by its unbiased estimator, and after applying mean-centering: $\shearmap := \ellmap - \langle \ellmap \rangle$, we get the inverse problem stated in \eqref{eq:invprob}, for which we seek an estimation of the convergence map $\convmap$.

\subsection{Masked data}
\label{subsec:appendix_pb_mask}

In practice, some data may be missing due to survey measurement masks (for instance, limits of the galaxy survey or bright stars in the foreground). We consider the set of ``active'' pixels, denoted by
\begin{equation}
    \MM := \set{k \in \smalloneto{\imgsize^2}}{\ngalperpix{k} > 0}.
\label{eq:activepixels}
\end{equation}
A naive approach consists in setting to $\shearmap[k] = 0$ for any $k \notin \MM$. Then, if we assign $\noise[k]$ to $-\convToShear\convmap[k]$ outside $\MM$, \eqref{eq:invprob} still holds, but $\noise$ can no longer be considered as a Gaussian noise. For the sake of simplicity, in this paper, we nevertheless relied on this solution, and assigned a very large noise variance $\covmatrNoise[k,\, k]$ for any $k \notin \MM$, which is required for the Wiener and MCALens mass mapping methods (see Sects.~\ref{subsec:appendix_mm_wien} and \ref{subsec:appendix_mm_mcal}, respectively).

Another approach involves
adding a Gaussian white noise $\noise'$ with large variance $\sigma'^2$ to the masked pixels: $\shearmap[k] := \noise'[k]$ for any $k \notin \MM$.
Assuming the S/N ratio between $\convToShear \convmap$ and $\noise'$ is small enough, we therefore approximately get \eqref{eq:invprob}, where the noise variance is left unchanged for unmasked pixels, and is equal to $\sigma'^2$ in masked regions.

Because of noise and masked data, this inverse problem is ill-posed. Therefore, mass-mapping methods generally rely on prior assumptions to regularize the problem.

\section{Mass mapping methods}
\label{sec:appendix_mm}

In this section, we present the three mass mapping methods used in our experiments.

\subsection{Kaiser-Squires inversion}

The Kaiser-Squires (KS) inversion \citep{Kaiser1993} consists in a simple pseudo-inversion of the linear operator $\convToShear$. According to \eqref{eq:shear2conv}, in absence of noise and mask, it provides an exact reconstruction of the true convergence map $\convmap$ (up to an additive constant). In practice, it is generally followed by a certain amount of Gaussian smoothing:
\begin{equation}
    \convmapEstimateKs := \BBS\convToShearPseudoinv \shearmap,
\label{eq:ks}
\end{equation}
where $\BBS \in \mathR^{\imgsize^2 \times \imgsize^2}$ encodes a Gaussian smoothing operator with manually-tuned variance.

Note that, in absence of smoothing ($\BBS = \BBI$), the KS solution corresponds to the maximum likelihood estimate of the convergence map:
\begin{align}
    \convmapEstimateKs
    &\in \argmin_{\convmap'} \Bigl\{
        -\log \condpdf{\cond{\shearmapRand}{\convmapRand}}{\shearmap}{\convmap'}
    \Bigr\} \\
    &= \argmin_{\convmap'} \frac12 \norm{\shearmap - \convToShear \convmap'}_{\covmatrNoise^{-1}}^2,
\label{eq:mle}
\end{align}
where $\condpdf{\cond{\shearmapRand}{\convmapRand}}{\shearmap}{\convmap'}$ denotes the likelihood density.

\subsection{Iterative Wiener filtering}
\label{subsec:appendix_mm_wien}

The KS inversion generally produces poor results because it is very sensitive to noise and masked data. To overcome this limitation, one introduce prior assumptions in the form of a regularization term in the objective function. Considering $\convmap$ as drawn from a random variable with probability density function $p$, the MAP estimate satisfies \eqref{eq:map}.

In the Wiener approach, we consider $\convmap$ as a Gaussian field. That is, it is drawn from a multivariate Gaussian distribution with (non-diagonal) covariance matrix $\covmatrConv$, and the covariance matrix of its Fourier transform $\tilde\convmap$, denoted by $\fouriercovmatrConv := \fouriermatr \covmatrConv \fouriermatrHerm$, is diagonal. Then, the distribution from which $\convmap$ is drawn is entirely characterized by its power spectrum. In this context, \eqref{eq:map} becomes
\begin{equation}
    \convmapEstimate_{\wiener} \in \argmin_{\convmap'} \left\{
        \frac12 \norm{\shearmap - \convToShear \convmap'}_{\covmatrNoise^{-1}}^2 + \frac12 \norm{\convmap'}_{\covmatrConv^{-1}}^2
    \right\}.
\end{equation}
By setting the gradient of the above expression to $0$, we get
\begin{equation}
    \convmapEstimate_{\wiener} = \left(
        \convToShear^{\herm} \covmatrNoise^{-1} \convToShear + \covmatrConv^{-1}
    \right)^{-1} \convToShear^{\herm} \covmatrNoise^{-1} \shearmap.
\label{eq:wien}
\end{equation}
If the noise is not stationary, then computing \eqref{eq:wien} requires two $\imgsize^2 \times \imgsize^2$-matrix inversions, which is not feasible in practice. To overcome this, \citet{Bobin2012} proposed an iterative algorithm converging toward \eqref{eq:wien}, in the context of cosmic microwave background (CMB) map restoration. Each iteration consists in a two-step process: one step in the spatial domain and one step in the Fourier domain. The algorithm takes advantage of the fact that the covariance matrices $\covmatrNoise$ and $\fouriercovmatrConv$ are diagonal, thereby avoiding large matrix inversions.

\subsection{MCALens algorithm}
\label{subsec:appendix_mm_mcal}

In contrast to the CMB, matter distribution is actually poorly approximated by Gaussian fields, due to the gravitational interactions responsible for the formation of galaxies and large-scale structures. In a more recent paper, \citet{Starck2021} suggested that convergence maps can be represented as the superposition of a Gaussian component, as in the Wiener solution, and a sparse component in a wavelet dictionary, as in GLIMPSE2D:
\begin{equation}
    \convmapEstimate_{\mcalens} = \convmapEstimate_{\gauss} + \convmapEstimate_{\sparse}.
\label{eq:mcalens}
\end{equation}
The authors then proposed to use the morphological component analysis (MCA) algorithm \citep{Starck2005}, considering that the Gaussian and sparse components are morphologically distinct (except for lower frequencies). They came along with an iterative algorithm called MCALens, where each iteration is a two-step process. First, perform one iteration of the GLIMPSE2D algorithm on the residual $\shearmap - \convToShear\convmapEstimate_{\gauss}$, and update the sparse component $\convmapEstimate_{\sparse}$. Then, perform one iteration of Wiener filtering, described in Sect.~\ref{subsec:appendix_mm_wien}, on the residual $\shearmap - \convToShear\convmapEstimate_{\sparse}$, and update the Gaussian component $\convmapEstimate_{\gauss}$.
Finally, since we consider the variations of the convergence around its mean value, the output of the MCALens algorithm undergoes mean-centering.

\section{Proof of Proposition~1}
\label{appendix:proofcqr}

\begin{proof}
    The exchangeability of $\calibrationsetRand{\sizeCalibrationset + 1}$ implies the exchangeability of the conformity scores $(\calibparamRand_i)_{i=1}^{\sizeCalibrationset + 1}$. Moreover, by hypothesis, the latter are almost surely distinct. Therefore, we can apply Lemma~2 by \citet[supplementary material]{Romano2019}:
    \begin{equation}
        1 - \alpha \leq \Proba\left\{
            \calibparamRand \leq \calibparamCQRRand{\alpha}
        \right\} \leq 1 - \alpha + \frac1{n+1},
    \label{eq:probaconformityscore}
    \end{equation}
    where we have denoted $\calibparamRand := \calibparamRand_{\sizeCalibrationset + 1}$.
    Let $\calibparam$ and $\calibparamCQR{\alpha}$ denote realizations of $\calibparamRand$ and $\calibparamCQRRand{\alpha}$, respectively. Then, $\calibparam$ corresponds to the (unknown) conformity score associated with the outcome $\inputOutput$ of $\inputOutputRand$. It satisfies, similarly to \eqref{eq:conformityscore1}-\eqref{eq:conformityscore2},
    \begin{align}
        \calibparam
            &= \min\left\{
                \calibparam' \in \interval{a}{b} :\, y \in \predintervCalib{\calibparam'}(x)
            \right\}
    \label{eq:conformityscore_unknown1} \\
            &= \min\left\{
                \calibparam' \in \interval{a}{b} :\, \calibfun_{\calibparam'}\bigl(\hat r(x)\bigr) \geq \left|
                    \hat f(x) - y
                \right|
            \right\}.
    \label{eq:conformityscore_unknown2}
    \end{align}
    Then, combining \eqref{eq:conformityscore_unknown1}, \eqref{eq:conformityscore_unknown2} and \eqref{eq:calibfun_nondecreasing} yields, for any $\calibparam' \in \mathR$,
    \begin{align}
        \calibparam \leq \calibparam' 
            &\iif \calibfun_{\calibparam'}\bigl(\hat r(x)\bigr) \geq \left|
                \hat f(x) - y
            \right| \\
            &\iif y \in \predintervCalib{\calibparam'}(x).
    \end{align}
    Therefore, using $\calibparam' := \calibparamCQR{\alpha}$, we get
    \begin{equation}
        \Proba\left\{
            \calibparamRand \leq \calibparamCQRRand{\alpha}
        \right\} = \Proba\left\{
            \SFY \in \predintervCalib{\calibparamCQRRand{\alpha}}(\SFX)
        \right\}.
    \label{eq:equivproba}
    \end{equation}
    Finally, plugging \eqref{eq:equivproba} into \eqref{eq:probaconformityscore} yields \eqref{eq:cqr_guarantee}, which concludes the proof.
\end{proof}

\onecolumn

\section{Detailed visualizations and results}
\label{sec:additional_fig_table}

Comprehensive visualizations of mass estimates and their uncertainties are shown in Fig.~\ref{fig:convestimates}, while detailed numerical results are provided in Table~\ref{table:results}.

\begin{table*}[h!]
    \setlength{\tabcolsep}{4pt}
    \caption{UQ metrics: miscoverage rate and mean length of prediction intervals}
    \label{table:results}
    \vspace{-5pt}
    \begin{centering}
    \small
    \begin{tabular}{ l l l | r r r r | r r r r }
        \hline\hline
        &&& \multicolumn{4}{c |}{Miscoverage rate ($\%$)}
        & \multicolumn{4}{c}{Mean length of pred.\@ intervals ($\times 10^{-2}$)} \\
        \multicolumn{3}{c |}{Type of calibration} & \multicolumn{1}{c}{KS1} & \multicolumn{1}{c}{KS2} & \multicolumn{1}{c}{Wiener} & \multicolumn{1}{c|}{MCALens} & \multicolumn{1}{c}{KS1} & \multicolumn{1}{c}{KS2} & \multicolumn{1}{c}{Wiener} & \multicolumn{1}{c}{MCALens} \\
        \hline
        \multicolumn{3}{l |}{Uncalibrated} & $6.94 \pm 0.39$ & $16.36 \pm 1.09$ & $36.04 \pm 1.93$ & $35.53 \pm 1.89$ & $11.43 \pm 0.00$ & $5.28 \pm 0.00$ & $2.65 \pm 0.02$ & $2.68 \pm 0.02$ \\
        \hline
        \multirow{7}{*}{CQR} & \multicolumn{2}{l |}{Add.\@} & $4.49 \pm 0.31$ & $4.53 \pm 0.53$ & $4.52 \pm 0.61$ & $4.52 \pm 0.55$ & $12.94 \pm 0.00$ & $8.33 \pm 0.00$ & $7.15 \pm 0.02$ & $\mathbf{7.13 \pm 0.02}$ \\
        & \multicolumn{2}{l |}{Mult.\@} & $4.49 \pm 0.31$ & $4.53 \pm 0.53$ & $4.53 \pm 0.60$ & $4.51 \pm 0.53$ & $12.94 \pm 0.00$ & $8.33 \pm 0.00$ & $7.39 \pm 0.04$ & $7.35 \pm 0.06$ \\
        & \multicolumn{2}{l |}{$\chi^2,\, a = 0.004$} & $4.49 \pm 0.31$ & $4.53 \pm 0.53$ & $4.53 \pm 0.60$ & $4.51 \pm 0.54$ & $12.94 \pm 0.00$ & $8.33 \pm 0.00$ & $7.31 \pm 0.04$ & $7.26 \pm 0.05$ \\
        & \multicolumn{2}{l |}{$\chi^2,\, a = 0.006$} & $4.49 \pm 0.31$ & $4.53 \pm 0.53$ & $4.53 \pm 0.60$ & $4.51 \pm 0.53$ & $12.94 \pm 0.00$ & $8.33 \pm 0.00$ & $7.38 \pm 0.04$ & $7.33 \pm 0.06$ \\
        & \multicolumn{2}{l |}{$\chi^2,\, a = 0.008$} & $4.49 \pm 0.31$ & $4.53 \pm 0.53$ & $4.53 \pm 0.60$ & $4.51 \pm 0.53$ & $12.94 \pm 0.00$ & $8.33 \pm 0.00$ & $7.42 \pm 0.05$ & $7.37 \pm 0.06$ \\
        & \multicolumn{2}{l |}{$\chi^2,\, a = 0.010$} & $4.49 \pm 0.31$ & $4.53 \pm 0.53$ & $4.53 \pm 0.60$ & $4.51 \pm 0.53$ & $12.94 \pm 0.00$ & $8.33 \pm 0.00$ & $7.45 \pm 0.05$ & $7.40 \pm 0.07$ \\
        & \multicolumn{2}{l |}{$\chi^2,\, a = 0.012$} & $4.49 \pm 0.31$ & $4.53 \pm 0.53$ & $4.53 \pm 0.60$ & $4.51 \pm 0.53$ & $12.94 \pm 0.00$ & $8.33 \pm 0.00$ & $7.48 \pm 0.05$ & $7.42 \pm 0.07$ \\
        \hline
        \multirow{6}{*}{RCPS} & \multirow{2}{*}{$\delta = 0.05$} & Add.\@ & $1.00 \pm 0.13$ & $1.05 \pm 0.18$ & $1.04 \pm 0.21$ & $1.04 \pm 0.17$ & $16.72 \pm 0.00$ & $12.45 \pm 0.00$ & $12.56 \pm 0.02$ & $12.26 \pm 0.02$ \\
        && Mult.\@ & $1.01 \pm 0.13$ & $1.05 \pm 0.18$ & $1.04 \pm 0.20$ & $1.03 \pm 0.16$ & $16.76 \pm 0.00$ & $12.61 \pm 0.00$ & $13.21 \pm 0.08$ & $12.75 \pm 0.12$ \\
        \cdashline{2-11}
        & \multirow{2}{*}{$\delta = 0.20$} & Add.\@ & $2.01 \pm 0.20$ & $2.05 \pm 0.30$ & $2.06 \pm 0.33$ & $2.06 \pm 0.30$ & $14.87 \pm 0.00$ & $10.05 \pm 0.00$ & $9.47 \pm 0.02$ & $9.35 \pm 0.02$ \\
        && Mult.\@ & $2.01 \pm 0.20$ & $2.05 \pm 0.30$ & $2.06 \pm 0.33$ & $2.05 \pm 0.28$ & $14.90 \pm 0.00$ & $10.15 \pm 0.00$ & $9.90 \pm 0.06$ & $9.71 \pm 0.09$ \\
        \cdashline{2-11}
        & \multirow{2}{*}{$\delta = 0.50$} & Add.\@ & $3.00 \pm 0.25$ & $3.06 \pm 0.41$ & $3.08 \pm 0.45$ & $3.07 \pm 0.41$ & $13.79 \pm 0.00$ & $8.96 \pm 0.00$ & $7.91 \pm 0.02$ & $7.88 \pm 0.02$ \\
        && Mult.\@ & $3.01 \pm 0.25$ & $3.06 \pm 0.40$ & $3.09 \pm 0.45$ & $3.06 \pm 0.39$ & $13.80 \pm 0.00$ & $9.03 \pm 0.00$ & $8.30 \pm 0.05$ & $8.23 \pm 0.07$ \\
        \hline
    \end{tabular}
    \end{centering}
    \vspace{5pt}

    {\textbf{Note.} The metrics are computed for each mass mapping method, before and after calibration, for various error levels $\delta$ (RCPS only), and various families of calibration functions $(\calibfun_\calibparam)_\calibparam$. ``Add.''\@ and ``Mult.''\@ refer to the families defined in \eqref{eq:calibfun_add} and \eqref{eq:calibfun_mult}, respectively, whereas ``$\chi^2$'' refers to the chi-squared family defined in \eqref{eq:calibfun_chisq}, with $k = 3$ degrees of freedom. The multiplicative factor $b$, which depends on the scaling factor $a$, is set to the maximum value such that $\calibfun_\calibparam$ remains non-decreasing for any $\calibparam \geq 0$. ``KS1'' and ``KS2'' correspond to the Kaiser-Squires estimators with weak and strong smoothing, respectively. The miscoverage rate \eqref{eq:miscoveragerate} as well as the length of the prediction intervals are averaged over the set $\MM$ of active pixels, for each pair $\pair{\shearmap_i}{\convmap_i}$. The empirical mean and standard deviation of each metric is then computed over the test set $\testsetMM$, and displayed in this table.}
\end{table*}

\begin{figure*}
    \centering
    \small
    \vspace{-8pt}
    \input{inputs/fig_convestimates}
    \vspace{-8pt}
    \caption{Top row: point estimate $\convmapEstimate$. Remaining rows: lower and upper bounds $\convmapEstimateLow$ and $\convmapEstimateHigh$, before and after calibration (additive calibration), for each of the three mass mapping methods. The bounds have been centered with respect to the ground truth $\convmap$.}
    \label{fig:convestimates}
\end{figure*}

\end{appendix}

\end{document}